\documentclass[preprint,twocolumn]{aastex63}

\usepackage{natbib}
\usepackage{enumerate}
\usepackage{graphicx}
\usepackage{epstopdf}
\usepackage{amsmath}
\usepackage{amssymb}
\usepackage{comment}
\usepackage{bm}
\usepackage{color}
\usepackage{multirow}
\usepackage{booktabs, makecell, multirow}
\usepackage{rotating}
\usepackage{hyperref}
\usepackage{bm}

\hypersetup{
    colorlinks=true,
    linkcolor=blue,
    urlcolor=blue,
}

\newcommand{\cref}{$^{(c)}$}

\newcommand{\mbh}{$M_{\rm BH}$}

\newcommand{\Hb}{H$\beta$}
\newcommand{\msigma}{$\rm{M}_{\rm{BH}}-\sigma_{*}$}
\newcommand{\logfmeansigma}{$\log_{10}(f_{{\rm mean},{\sigma}})$}
\newcommand{\logfrmssigma}{$\log_{10}(f_{{\rm rms},{\sigma}})$}
\newcommand{\logfrmsfwhm}{$\log_{10}(f_{{\rm rms},{\rm FWHM}})$}
\newcommand{\logfmeanfwhm}{$\log_{10}(f_{{\rm mean},{\rm FWHM}})$}
\newcommand{\lineprofilerms}{$\log_{10}(\rm{FWHM}/\sigma)_{\rm{rms}}$}
\newcommand{\lineprofilemean}{$\log_{10}(\rm{FWHM}/\sigma)_{\rm{mean}}$}
\newcommand{\lineprofile}{$\log_{10}(\rm{FWHM}/\sigma)$}

\newcommand{\three}{3C 382}

\defcitealias{pancoast14b}{P14}
\defcitealias{Grier++17}{G17}
\defcitealias{2018ApJ...866...75W}{W18}
\defcitealias{2020ApJ...902...74W}{W20}
\defcitealias{bentz2021detailed}{B21}
\defcitealias{pancoast11}{P11}
\defcitealias{Villafana22}{V22}
\defcitealias{2023ApJ...948...95V}{V23}
\defcitealias{2022ApJ...934..168B}{B22}
\defcitealias{2023ApJ...944...29B}{B23b}
\defcitealias{2023ApJ...959...25B}{B23a}
\defcitealias{Wang_2026}{W26}

\shorttitle{Can BLR line profile shape improve single-epoch black hole mass estimates? }
\shortauthors{Villafa\~na et al.}
 
\begin{document}
\title{Can BLR line profile shape improve single-epoch black hole mass estimates? }
\author[0000-0002-1961-6361]{Lizvette Villafa\~na}
\affiliation{Physics Department, California Polytechnic State University, San Luis Obispo CA 93407, USA}

\author[0000-0002-8460-0390]{Tommaso Treu}
\affiliation{Department of Physics and Astronomy, University of California, Los Angeles, CA 90095-1547, USA}

\author[0000-0002-2052-6400]{Shu Wang}
\affiliation{Astronomy Program, Department of Physics and Astronomy, Seoul National University, Seoul, 08826, Republic of Korea}

\author[0000-0002-2816-5398]{Misty C. Bentz}
\affiliation{Department of Physics and Astronomy, Georgia State University, Atlanta, GA 30303, USA}

\author[0000-0001-9902-7112]{Brendon J. Brewer}
\affiliation{Department of Statistics, The University of Auckland, Private Bag 92019, Auckland 1142, New Zealand}

\author[0000-0002-3026-0562]{Aaron J. Barth}
\affiliation{Department of Physics and Astronomy, 4129 Frederick Reines Hall, University of California, Irvine, CA 92697, USA}

\author[0000-0002-8055-5465]{Jong-Hak Woo}
\affiliation{Astronomy Program, Department of Physics and Astronomy, Seoul National University, Seoul, 08826, Republic of Korea}

\author[0000-0001-6919-1237]{Matthew A. Malkan}
\affiliation{Department of Physics and Astronomy, University of California, Los Angeles, CA 90095-1547, USA}

\author[0000-0003-2064-0518]{Vardha N. Bennert}
\affiliation{Physics Department, California Polytechnic State University, San Luis Obispo CA 93407, USA}

\author[0000-0002-1912-0024]{Vivian U}
\affiliation{Caltech/IPAC, 1200 E. California Blvd., Pasadena, CA 91125}

\correspondingauthor{Lizvette Villafa\~na}
\email{lvillafa@calpoly.edu}

\begin{abstract}
The virial coefficient ($f$), which is meant to encapsulate broad-line region (BLR) geometry and kinematics, remains one of the largest sources of systematic uncertainty in black hole mass estimates for Active Galactic Nuclei (AGNs). While the use of a sample average $\langle f \rangle$ enables black hole mass estimates across large samples and cosmological distances, individual AGNs may deviate from this average due to differences in BLR structure and viewing angle. In previous work, we reported marginal evidence for a correlation between $f$ and the shape of the broad H$\beta$ emission line, $\log_{10}(\mathrm{FWHM}/\sigma)$.
In this work, we update our sample to include ten new sources with \texttt{CARAMEL} BLR dynamical modeling, increasing both the black hole mass range and statistical power of our analysis. We find marginal evidence for a correlation between $f$ and $\log_{10}(\mathrm{FWHM}/\sigma)$, with a slope and intrinsic scatter consistent with previous results. The confirmation of this trend across a larger sample further supports the idea that line profile shape may reflect BLR properties in a way that directly impacts $f$. If confirmed with future BLR dynamical modeling of sources within a wider range of $\log_{10}(\mathrm{FWHM}/\sigma)$, this relationship could enable empirical estimates of the virial coefficient and improve single-epoch black hole mass estimates across cosmic time.
\keywords{Seyfert galaxies, active galaxies, supermassive black holes, reverberation mapping}
\end{abstract}

\section{Introduction}\label{sec:intro}
It is widely accepted that a supermassive black hole (SMBH) lies in the center of nearly every galaxy. Despite their complexity, SMBHs are fundamentally described by just three parameters: mass, spin, and charge, with mass being the most accessible observationally. Precise black hole mass measurements are therefore essential for constructing reliable black hole demographics across cosmic time, as well as advancing our understanding of SMBH seeds \citep[e.g.,][]{2010A&ARv..18..279V, 2020ARA&A..58...27I} and their role in galaxy evolution \citep[e.g.][]{Ferrarese00,Gebhardt00, Gultekin09,Woo10,Bennert11,Bennert15,McConnell_Ma13}. 

Many direct methods for determining the mass of a black hole involve modeling the motions of gas or stars within its gravitational sphere of influence \citep[e.g.,][]{kormendy95,ferrarese05}. This approach, however, is limited to nearby galaxies where such regions can be spatially resolved; at larger distances, this becomes increasingly difficult. To overcome this limitation, broad line region (BLR) reverberation mapping utilizes the variability of Type 1 Active Galactic Nuclei (AGNs) to resolve the gravitational sphere of influence in time, enabling \mbh\ measurements beyond the local universe (\citealt{blandford82}; \citealt{peterson93}; for a review, see \citealt{2021iSci...24j2557C}). 

In Type 1 AGN, the time delay ($\tau$) measured between variations in the continuum and broad emission lines provides the radius of the BLR. Assuming the BLR is virialized, the velocity ($v$) of the gas is measured from the width of the broad emission line. Combining these two measurements yields \mbh\ via the virial relation: 
\begin{equation}
    M_{\rm BH} = f \frac{c\tau v^2}{G} = fM_{\rm{vir}},
    \label{eq: rev_map}
\end{equation}
where $G$ is the gravitational constant, $c$ is the speed of light,  and the virial product is defined as  $M_{\rm{vir}}=c\tau v^2/G$. The dimensionless scale factor $f$, often referred to as the ``virial coefficient," is a factor of order unity and depends on the structure, kinematics, and orientation of the BLR \citep{peterson04}.

Since reverberation mapping is observationally expensive and challenging, especially at high redshift, most \mbh\ are estimated using the single epoch method. Using a single spectrum, the empirical BLR size-luminosity (R-L) relation replaces $c\tau$ in Eqn. \ref{eq: rev_map} with an estimate of $r_{\rm{BLR}}$ based on the observed correlation between nuclear luminosity and BLR size, established by the reverberation mapped AGN sample \citep{1999ApJ...526..579W,2000ApJ...533..631K,2009ApJ...697..160B, bentz13, Grier17b, Du19,Fonseca-Alvarez20, Shen24, Woo24, Wang24}. These single epoch estimates still require a virial coeffient, $f$, and therefore carry the same systematic uncertainties related to BLR geometry and kinematics as reverberation mapping based estimates.

Generally, a sample average virial coefficient $\langle f \rangle$, found by forcing the sample of reverberation mapped AGNs to align with the \msigma\ relation of local quiescent galaxies, is adopted \citep[e.g.,][]{Onken04,2006A&A...456...75C,Woo10,Woo13,Woo15,2012ApJ...747...30P,Grier13b,Hokim14,Hokim15,Batiste17}. While this approach has enabled \mbh\ estimates across cosmic time, the virial coefficient is expected to vary among individual AGNs due to differences in physical properties. For example, the virial coefficient is expected to depend on inclination, as projection effects influence the observed line-of-sight velocity, e.g., a thin disk BLR viewed at low inclination would lead to an underestimation of \mbh\ \citep{2001ApJ...551...72K}. However, without the ability to measure $f$ directly for individual sources, capturing and understanding systematic trends in the virial coefficient (e.g., due to inclination) has been limited. Several studies have instead investigated systematic trends in \mbh\ estimates  \citep[e.g.,][]{2006A&A...456...75C, Meja-Restrepo18, Yu19}. Yet, due to the unresolved nature of the BLR and inability to measure $f$ directly, it remains uncertain which BLR/AGN physical properties beyond inclination influence the virial coefficient and how they translate into $f$ \citep[see][for recent interferometric constraints]{Gravity18, Gravity20, Gravity21}. 

The search for such empirical trends in the virial coefficient has been ongoing since the earliest BLR dynamical modeling samples \citep[e.g.,][]{pancoast14a,Grier++17,2018ApJ...866...75W,Villafana22,2023ApJ...948...95V}, leveraging the unique ability of this approach to constrain individual AGN-specific virial coefficients. BLR dynamical modeling directly infers the virial coefficient for individual AGNs by forward modeling velocity-resolved RM data sets, following the framework first introduced by \citet{pancoast11} and implemented in the Code for AGN Reverberation and Modeling of Emission Lines (\texttt{CARAMEL}). Using a Bayesian framework, \texttt{CARAMEL} explores a phenomenological description of the BLR and constrains
a black hole mass that is consistent with the velocity-resolved reverberation mapping dataset, independent of any assumed $\langle f \rangle$. Combining the derived \texttt{CARAMEL} black hole mass with the virial product ($M_{\rm{vir}}$), enables AGN-specific estimates of the virial coefficient. 

Recent studies have leveraged BLR dynamical modeling results to derive a (BLR dynamical modeling) sample average virial coefficient $\langle f \rangle$ independent of the traditional \msigma\ calibration \citep[e.g.,][]{Shen24, Winkel25}. Using the full \texttt{CARAMEL} BLR dynamical modeling sample compiled in this work, \textcolor{blue}{\citet{Wang_2026}} determine an average $f$-value consistent with traditional \msigma\ calibrations, but with a notably smaller intrinsic scatter $(\sim 0.2 \rm{~dex})$. This result demonstrates the power of using BLR dynamical modeling  
to help refine \mbh\ measurements -- though an average $\langle f \rangle$ is still required. In this paper, we take a complementary approach: rather than focusing on calibrating an average $\langle f \rangle$, we explore whether $f$ varies systematically with observable BLR or AGN properties. 

Previous work, \citet[][hereafter \citetalias{2023ApJ...948...95V}]{2023ApJ...948...95V}, analyzed 28 AGN with BLR dynamical modeling and found marginal evidence for a correlation between the virial coefficient $f$ and line profile shape $\log_{10}(\mathrm{FWHM}/\sigma)$. This trend is physically motivated, as the shape of the broad \Hb\ line encodes structural and kinematic properties of the BLR \citep[eg., ][]{2006A&A...456...75C, 2011BaltA..20..400K, 2013A&A...549A.100K}, therefore making \lineprofile\ a plausible tracer of the virial coefficient. If confirmed with a larger and more diverse sample, a \lineprofile-$\log_{10}f$ scaling relation could offer a practical path toward improving single-epoch \mbh\ estimates without assuming a universal virial coefficient, $\langle f \rangle$. 

Since our previous \citetalias{2023ApJ...948...95V} work, there have been 10 additional sources with \textsc{caramel} BLR dynamical modeling. We extend the sample presented in \citetalias{2023ApJ...948...95V} to 38 sources by incorporating one AGN from \citet[][hereafter \citetalias{2023ApJ...959...25B}]{2023ApJ...959...25B}, one from  \citet[][hereafter \citetalias{2023ApJ...944...29B}]{2023ApJ...944...29B}, and 8 additional AGN from the Seoul National University AGN Monitoring Project \citep[SAMP;][hereafter \citetalias{Wang_2026}]{Wang_2026}. The multi-year ($\sim$ 6 years) SAMP reverberation mapping campaign targeted high luminosity AGNs \citep{Woo19, Rakshit19, Cho23, Woo24, Wang25}. The sample is thus more representative of high-redshift AGNs, which the single-epoch method is used to estimate \mbh\ \citep{Woo24}. Using this updated sample, we search for correlations between the \Hb\ virial coefficient and BLR parameters, including inclination, Eddington ratio, and line profile shape. In particular, the addition of the SAMP sources significantly expands the luminosity regime of the \textsc{caramel} sample to $\log_{10}(L_{5100}/\rm{erg\ s}^{-1}) \sim 44-45$. This expanded sample provides an important test of the potential empirical $f$-\lineprofile\ relation by increasing the statistical power of our analysis and allowing us to test whether this relation remains consistent across a larger sample, particularly in the higher-luminosity regime relevant for the single-epoch method.

This paper is organized as follows: in Section \ref{sec: caramel}, we briefly summarize the \texttt{CARAMEL} BLR dynamical model. In Section \ref{sec: sample}, we describe the most recent compilation of sources with \texttt{CARAMEL} BLR dynamical modeling. In Section \ref{sec:RESULTS}, we present the results of correlation tests between the \Hb\ virial coefficient and various BLR and AGN properties. We discuss the implications of these trends in Section \ref{sec: discussion}, and summarize our conclusions in Section \ref{sec: summary}.

\section{Summary of BLR Dynamical Model} \label{sec: caramel}
We employ the BLR dynamical modeling methods first introduced by \citet{pancoast11}. In particular, our sample (outlined in Section \ref{sec: sample}) includes all sources with BLR dynamical modeling results produced using the Code for AGN Reverberation and Modeling of Emission Lines (\texttt{CARAMEL}). Here we briefly summarize the \texttt{CARAMEL} code and relevant model parameters used in this work.

Within a Bayesian framework, \texttt{CARAMEL} explores a phenomenological description of the BLR \textit{emission} that is consistent with the velocity-resolved reverberation mapping data set and constrains \mbh\ without the need of assuming a virial coefficient. The \texttt{CARAMEL} code explores the BLR emissivity distribution of a single emission line by modeling the BLR as point particles that surround the central black hole and instantaneously re-emit light received from the ionizing source toward the observer \citep[see][for recent \texttt{CARAMEL-GAS} efforts to model the underlying gas distribution, rather than emissivity distribution]{2022ApJ...935..128W, 2024ApJ...966..106V}.

The radial distribution of the BLR point particles is drawn from a gamma distribution with shape parameter $\alpha$ and scale parameter $\theta$,
\begin{equation}
    p(x|\alpha,\theta)\propto x^{\alpha-1}\exp\Big(-\frac{x}{\theta}\Big)\,
\end{equation}
\noindent shifted from the origin by the Schwarzschild radius (plus a free parameter, $r_{\mathrm{min}}$, which sets the minimum BLR radius). Once the radial distribution of particles is set, the opening angle $\theta_o$ determines the disk thickness, with values of $\theta_o \rightarrow 0^\circ $ representing a razor thin disk and $\theta_o \rightarrow 90^\circ $ representing a spherical BLR. Then, the inclination angle, $\theta_i$ is defined as the angle between the BLR disk and the observer's line of sight, with values of $\theta_o \rightarrow 0^\circ $ representing a face-on geometry and $\theta_o \rightarrow 90^\circ $ representing an edge-on geometry. 

While the location of the BLR point particles determines the particle's associated time lag, the particle's line-of-sight velocity determines the corresponding shift in wavelength of the modeled emission. The velocity distribution of the particles is determined by a number of parameters: the $f_{\rm{ellip}}$ parameter determines the fraction of particles with nearly circular Keplerian orbits and the remaining particles exhibit inflowing/outflowing behavior as determined by the binary parameter $f_{\rm{flow}}$. Additionally, the model parameter $\sigma_{\rm{turb}}$ allows for macroturbulent velocities. Additional model parameters allow for asymmetries in the BLR emission, but these parameters are not discussed in this work and we refer the reader to \citet{pancoast14b} for a complete description of the \texttt{CARAMEL} model.

Lastly, we note that as mentioned in \citetalias{2023ApJ...948...95V}, some minor changes were made to the \textsc{caramel} code since its original publication. As shown in \citet{2024ApJ...966..106V}, we have found that the updated code used by \citet{Villafana22} and \citet{Wang_2026} does not significantly change the results produced by the original code \citep[e.g.,][]{2018ApJ...866...75W, Grier++17, pancoast14a, 2020ApJ...902...74W, bentz2021detailed, 2022ApJ...934..168B,  2023ApJ...959...25B, 2023ApJ...944...29B}. All future \textsc{caramel} work will be based on the updated version (updates outlined in \citetalias{2023ApJ...948...95V} Appendix).

\begin{deluxetable}{llcccc}
\setlength{\tabcolsep}{9pt}
\tablecaption{Sources with CARAMEL modeling}
\tablehead{
\colhead{Galaxy} & 
\colhead{Redshift} & 
\colhead{Ref.}}
\startdata
            $\rm 3C~120$ &  0.0330 &            \citetalias{Grier++17} \\
           $\rm Arp~151$ &  0.0211 &          \citetalias{pancoast14b} \\
          $\rm IC~4329A$ &  0.0160 &  \citetalias{2023ApJ...959...25B} \\
         $\rm J0140+234$ &  0.3200 &                               \citetalias{Wang_2026} \\
         $\rm J1026+523$ &  0.2590 &                               \citetalias{Wang_2026} \\
         $\rm J1120+423$ &  0.2260 &                              \citetalias{Wang_2026}\\
         $\rm J1217+333$ &  0.1780 &                               \citetalias{Wang_2026} \\
         $\rm J1540+355$ &  0.1640 &                                \citetalias{Wang_2026} \\
      $\rm MCG+04-22-04$ &  0.0323 &       \citetalias{Villafana22} \\
        $\rm Mrk ~ 1048$ &  0.0431 &       \citetalias{Villafana22} \\
        $\rm Mrk ~ 1392$ &  0.0361 &       \citetalias{Villafana22} \\
         $\rm Mrk ~ 841$ &  0.0364 &       \citetalias{Villafana22} \\
          $\rm Mrk~1310$ &  0.0194 &          \citetalias{pancoast14b} \\
           $\rm Mrk~141$ &  0.0417 &  \citetalias{2018ApJ...866...75W} \\
          $\rm Mrk~1501$ &  0.0893 &            \citetalias{Grier++17} \\
          $\rm Mrk~1511$ &  0.0339 &  \citetalias{2018ApJ...866...75W} \\
           $\rm Mrk~279$ &  0.0305 &  \citetalias{2018ApJ...866...75W} \\
           $\rm Mrk~335$ &  0.0258 &            \citetalias{Grier++17} \\
            $\rm Mrk~50$ &  0.0234 &  \citetalias{2018ApJ...866...75W} \\
          $\rm NGC~3227$ &  0.0038 &  \citetalias{2023ApJ...944...29B} \\
          $\rm NGC~3783$ &  0.0097 &    \citetalias{bentz2021detailed} \\
          $\rm NGC~4151$ &  0.0033 &  \citetalias{2022ApJ...934..168B} \\
          $\rm NGC~4593$ &  0.0090 &  \citetalias{2018ApJ...866...75W} \\
          $\rm NGC~5548$ &  0.0172 &          \citetalias{pancoast14b} \\
  $\rm NGC~5548~(STORM)$ &  0.0172 &  \citetalias{2020ApJ...902...74W} \\
          $\rm NGC~6814$ &  0.0052 &          \citetalias{pancoast14b} \\
     $\rm NPM1G+27.0587$ &  0.0620 &       \citetalias{Villafana22} \\
     $\rm PG ~ 2209+184$ &  0.0700 &       \citetalias{Villafana22} \\
       $\rm PG~0947+396$ &  0.2060 &                                \citetalias{Wang_2026} \\
       $\rm PG~1121+422$ &  0.2250 &                              \citetalias{Wang_2026}\\
       $\rm PG~1310-108$ &  0.0343 &  \citetalias{2018ApJ...866...75W} \\
       $\rm PG~1427+480$ &  0.2200 &                               \citetalias{Wang_2026} \\
       $\rm PG~2130+099$ &  0.0630 &            \citetalias{Grier++17} \\
        $\rm RBS ~ 1303$ &  0.0418 &       \citetalias{Villafana22} \\
        $\rm RBS ~ 1917$ &  0.0660 &       \citetalias{Villafana22} \\
 $\rm RXJ ~ 2044.0+2833$ &  0.0500 &       \citetalias{Villafana22} \\
          $\rm SBS~1116$ &  0.0279 &          \citetalias{pancoast14b} \\
            $\rm Zw~229$ &  0.0279 &  \citetalias{2018ApJ...866...75W} \\
       \hline 
\enddata
\tablecomments{Full sample of sources with \texttt{CARAMEL} BLR dynamical modeling to date. Galaxy name and redshift are found in Columns 1 and 2, respectively. Column 3 refers to the published \textsc{caramel} results. The sample has increased from 28 to 38, with the additional sources of \citetalias{2023ApJ...959...25B}, \citetalias{2023ApJ...944...29B}, \citetalias{Wang_2026} } 
\label{tab: extended_sample}
\end{deluxetable} 

\section{The \texttt{CARAMEL} Sample} \label{sec: sample}
We compile the most up-to-date sample of AGN with \texttt{CARAMEL} BLR dynamical modeling (see Table~\ref{tab: extended_sample}) to investigate correlations between the inferred H$\beta$ virial coefficient and AGN/BLR properties. In addition to the 28 AGN analyzed in \citetalias{2023ApJ...948...95V}, we also include recent \texttt{CARAMEL} results for IC 4329A and NGC 3227 \citepalias{2023ApJ...959...25B, 2023ApJ...944...29B}, and 8 AGN from the Seoul National University AGN Monitoring Project \citepalias[SAMP;][]{Wang_2026}, bringing the total 38 \texttt{CARAMEL} sources.\footnote{We note that NGC 5548 appears twice in the sample, reflecting independent \texttt{CARAMEL} modeling of the LAMP 2008 and AGNSTORM campaigns.} 

In particular, the inclusion of the SAMP sample extends the luminosity range of our dataset to $\log_{10}(L_{5100}/\rm{erg\ s}^{-1}) \sim 44-45$, and correspondingly the black hole mass range to $M_{\rm BH} \sim 10^8$--$10^9\,M_\odot$, allowing us to search for potential trends across a broader population. 

Although a larger number of AGNs have BLR dynamical modeling results using other codes --- such as \texttt{BRAINS} \citep{li13,Li2018, Stone24} and the \textsc{agn-gravity} collaboration's framework \citep{Gravity18,Gravity20,Gravity21,Gravity24}
--- we restrict our analysis to the sample compiled using only the \texttt{CARAMEL} modeling code. This minimizes potential systematic uncertainties due to differences in code implementation, even though the underlying modeling frameworks are similar. Additionally, we focus exclusively on the H$\beta$-emitting BLR, although \texttt{CARAMEL} has been applied to other emission lines (e.g., see \citealt{2020ApJ...902...74W} for Ly$\alpha$ and \ion{C}{4} in NGC 5548; and \citealt{bentz2021detailed} for \ion{He}{2} in NGC 3783),  to ensure further consistency across our sample. 

In the sections below, we summarize how the AGN-specific virial coefficients are inferred (Section \ref{sec: f}) and list the AGN/BLR parameters used in our analysis (Section \ref{sec: parameters}). 
\subsection{Sample AGN specific virial coefficients} \label{sec: f}
In addition to learning about the BLR structure and kinematics, BLR dynamical modeling is of particular interest for its ability to constrain \mbh\ without the need of invoking the virial coefficient. Importantly, BLR dynamical modeling provides the only way to infer an individual AGN-specific virial coefficient. 

The virial coefficient of an AGN is determined post-process, using the resulting \mbh\ posterior distribution function (PDF) from \texttt{CARAMEL}. Summarizing the method outlined in \citetalias{2023ApJ...948...95V}, from Eqn.\ \ref{eq: rev_map}, it follows that the AGN-specific virial coefficient can be inferred using the \texttt{CARAMEL} \mbh\ result and the virial product $M_{\rm{vir}}$ inferred from reverberation mapping:
\begin{equation}
    f_{\rm{spec\ type,\ line\ width}} = \frac{M_{\rm{BH}} G}{c\tau v^2} = \frac{M_{\rm{BH}}}{M_{\rm{vir}}},
    \label{eq: agn-f}
\end{equation}
where ``spec type'' and ``line width'' represent the type of spectra and definition of line width, $v$, used to determine the virial product ($M_{\rm{vir}}$), respectively (e.g., FWHM or $\sigma_{\rm{line}}$ as measured from the mean or rms spectrum),  $\tau$ is the cross correlation time-lag measured from traditional reverberation mapping analyses, and $M_{\rm{BH}}$ represents the \texttt{CARAMEL} \mbh\ result. 

To propagate uncertainty, we assume Gaussian errors on the cross-correlation time-lag and line width measurements used to determine the virial product ($M_{\rm{vir}}=c\tau v^2/G$). For measurements reported with asymmetrical uncertainties, we adopt the mean of the measurement's upper and lower uncertainties. The resulting AGN-specific virial coefficient is also a distribution function, and we use the 68\% confidence interval for 1$\sigma$ uncertainties when reporting the inferred virial coefficient for an individual AGN. We note that use of the rms spectrum and line dispersion have been found to produce less biased results \citep{peterson04, 2006A&A...456...75C, Wang19, Dalla-Bonta20}, however the FWHM is often used if the line dispersion measurement is unavailable. For completeness, we calculate four different virial coefficients per source: using either the rms and mean spectra, and either the FWHM or line dispersion $\sigma_{\rm{line}}$ line widths -- i.e., $f_{\rm{rms,}\sigma}$, $f_{\rm{rms, FWHM}}$, $f_{\rm{mean,}\sigma}$, $f_{\rm{mean, FWHM}}$. 

\subsection{Sample AGN/BLR parameters}
\label{sec: parameters}
In this work, we search for trends between the inferred AGN-specific virial coefficient and AGN/BLR parameters, as well as line profile shape, i.e -- the ratio of line widths, $\log_{10}(\rm{FWHM}/\sigma)$. Tables with the relevant parameters of our sample, listed below, are found in the Appendix:
\begin{enumerate}
    \item Inferred \texttt{CARAMEL} \mbh\ estimates and inferred AGN-specific virial coefficients (Table \ref{tab:table_individual_f}). 
    \item Line widths measurements with corresponding references (Table \ref{tab: line_widths}).
    \item Inferred \texttt{CARAMEL} BLR inclination and opening angles (Table \ref{tab: geometry}).
\end{enumerate}
All other \texttt{CARAMEL} modeling results of a source can be found in the corresponding \texttt{CARAMEL} paper listed in Table \ref{tab: extended_sample}.

\subsubsection{Bolometric Correction and the Eddington Ratio}
To estimate Eddington ratios, $\log_{10}(L_{\rm{bol}}/L_{\rm{Edd}})$, we \textbf{use }a fixed bolometric correction factor of 9, consistent with previous studies \citep[e.g., ][and based on a spectral energy distribution with a strong blue bump]{2000ApJ...533..631K,peterson04,2013BASI...41...61S}. We apply the bolometric correction factor to the extinction-corrected monochromatic luminosity at 5100$\mathrm{\AA}$ (with host-galaxy starlight removed), such that $L_{bol}\approx 9 \lambda L_{\lambda}(5100 \mathrm{\AA})$. While bolometric corrections may in reality depend on the Eddington ratio itself, and {$L_{bol}\approx 9 \lambda L_{\lambda}(5100 \mathrm{\AA})$} may not be appropriate for all types of AGNS \citep{peterson04}, we adopt a factor of 9 to remain consistent with prior \texttt{CARAMEL} studies \citep[e.g.,][]{pancoast14b, Grier++17, 2018ApJ...866...75W, Villafana22}. 

While $L_{bol}\approx 9 \lambda L_{\lambda}(5100 \mathrm{\AA})$ is a useful approximation for determining the bolometric luminosity, we emphasize that trends with Eddington ratio should be interpreted in the context of this assumption. Additionally, one should also consider the systematic uncertainties involved with possible host-galaxy starlight contamination. For most sources, the optical $\lambda L_{\lambda}(5100 \mathrm{\AA})$ luminosities were measured using spectral decompositions. In these fits, each spectrum is modeled as a combination of an AGN power-law continuum, \ion{Fe}{2} emission, and a host-galaxy stellar component \citep[for details regarding the spectral decomposition, see][]{barth11, Woo24}. This allows the host starlight contribution at $\lambda L_{\lambda}(5100 \mathrm{\AA})$ to be approximately subtracted, however, the luminosities and Eddington ratios may be subject to overestimation, particularly for the \citetalias{2023ApJ...948...95V} sample. For the high-luminosity SAMP sources, host galaxy contamination is not expected to be significant given the AGN-dominated flux \citep[average host fraction of 0.07;][]{Woo24}. For NGC 3227 \citep{2023ApJ...959...25B} and IC 4329A \citep{2023ApJ...944...29B}, the host galaxy contribution was determined using HST images and the methods of \citet{2009ApJ...697..160B, bentz13}. 

\subsection{Model Limitations}
While the previous \texttt{CARAMEL} sample of 28 AGNs primarily included moderate-luminosity sources \citep{Villafana22},  the current sample incorporates SAMP targets, which probe a higher-luminosity regime by $\sim 1$ dex. In particular, the updated sample probes a slightly broader Eddington ratio, but does not substantially probe the super-Eddington regime (see Figure \ref{fig: thetaovseddington}). In fact, of the 38 sources in our sample, only one source approaches $\log_{10}(L_{\rm{bol}}/L_{\rm{Edd}}) \sim 0.5$.
This distinction is important when considering model limitations \citep[see][for a detailed discussion]{2020MNRAS.493.1227R}.  
In its current implementation, \texttt{CARAMEL} only accounts for gravitational effects from the central black hole and does not include contributions from radiation pressure. While radiation pressure may become increasingly important near the Eddington limit, the predominately sub-Eddington nature of our sample of 38 AGNs suggests gravity alone should provide a reasonable description of the BLR. At the same time, models such as the Failed Radiatively Accelerated Dusty Outflow (\texttt{FRADO}) framework \citep[e.g.,][]{2017ApJ...846..154C} indicate that radiation pressure can influence BLR structure even in the sub-Eddington regime, highlighting the value of further development of BLR dynamical modeling approaches to incorporate radiation pressure effects, which is beyond the scope of this paper. 

Additionally, we note that there is a known modeling degeneracy between inclination and opening angles \citep[see][for further discussion]{Grier++17, 2026ApJ..1000...48B}. However, \citet{2026ApJ..1000...48B} recently provided independent evidence of the robustness of inferred \textsc{caramel} inclination angles by comparison to jet inclination and accretion disk measurements.

  \begin{figure*}[ht]
    \centering
    \begin{minipage}[b]{1.0\textwidth}
        \centering
        \includegraphics[width=1.0\textwidth]{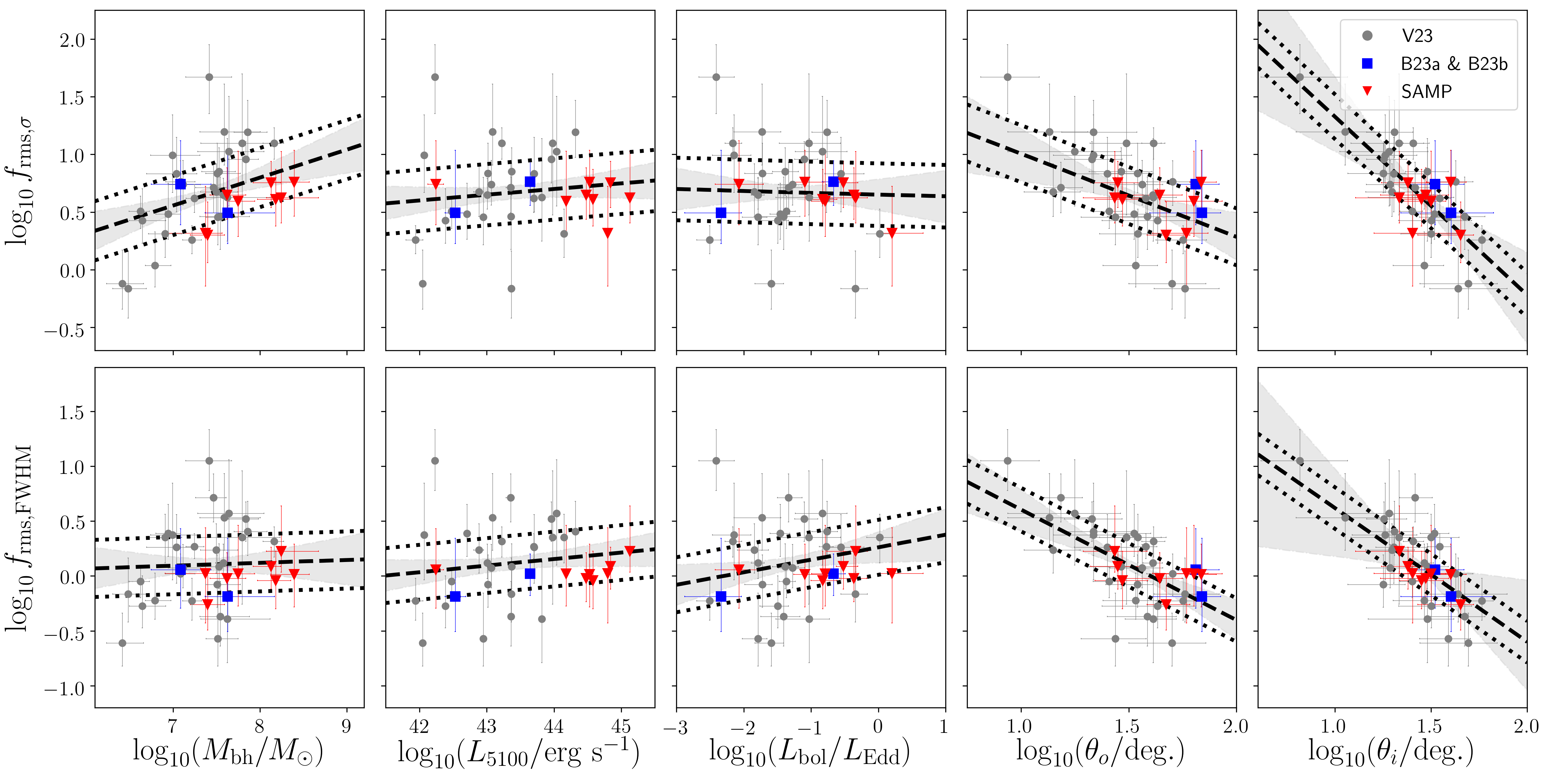}
        \caption{Correlations between \logfrmssigma\ and \logfrmsfwhm\ with select BLR/AGN parameters. From left to right: \mbh, optical luminosity, Eddington ratio, \Hb-emitting BLR opening angle (disk thickness), and \Hb-emitting BLR inclination angle. The dashed black lines and gray shaded regions give the median and 68\% confidence intervals of the linear regression. Dotted lines are offset above and below the dashed line by the median value of the intrinsic scatter. Grey points represent the  sample used in \citetalias{2023ApJ...948...95V}. This work includes the addition of 10 new sources -- \citetalias{2023ApJ...959...25B} and \citetalias{2023ApJ...944...29B} are indicated by blue squares, and SAMP \citetalias{Wang_2026} are shown in red triangles. Note the error bars plotted are based on the 68\% confidence interval of the model parameters. Covariance between model parameters is taken into account in the linear regression, but for readers interested in the 2D posterior PDFs, please refer to Figure 
        found in the Appendix.}
        \label{fig:f_correlations_rms}
    \end{minipage}
    
    \vspace{0.1in}  
    
    \begin{minipage}[b]{1.0\textwidth}
        \centering
        \includegraphics[width=\textwidth]{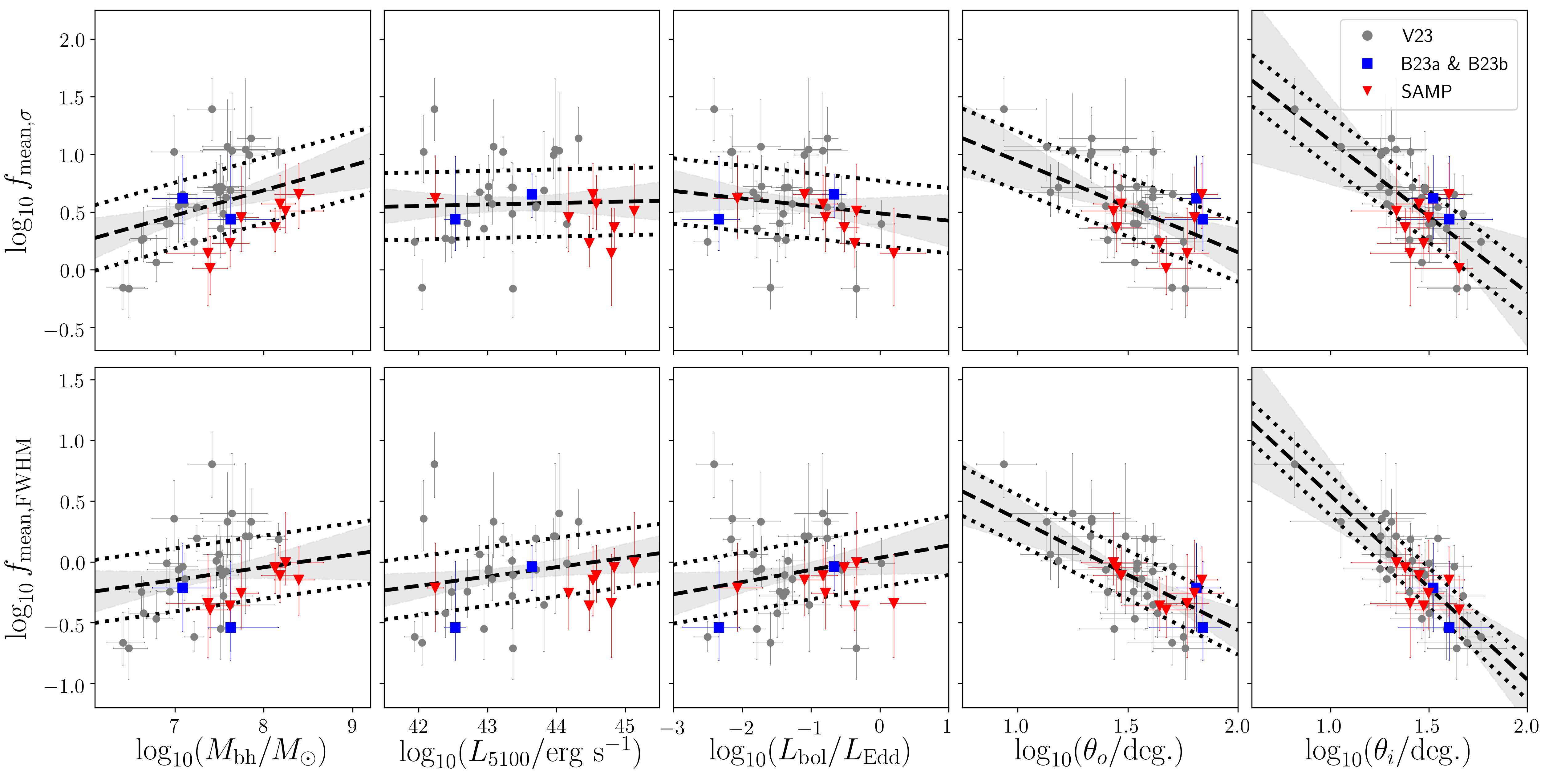}
        \caption{Same as Figure \ref{fig:f_correlations_rms}, for correlations between \logfmeansigma\ and \logfmeanfwhm\ with select BLR/AGN parameters. For readers interested in the 2D posterior PDFs, please refer to Figure 
        found in the Appendix.}
        \label{fig:f_correlations_mean}
    \end{minipage}
\end{figure*}

\section{Results}\label{sec:RESULTS}
To search for trends with the inferred AGN-specific \Hb\ virial coefficients and AGN/BLR parameters, we perform a Bayesian linear regression using the \textsc{IDL} routine \texttt{linmix\_err} \citep{Kelly07}. Since the AGN-specific virial coefficients are determined using the \texttt{CARAMEL} \mbh\ estimates, we must account for correlated uncertainties. Using a Bayesian linear regression isolates intrinsic correlations and avoids capturing artifacts of shared measurement errors. We note that while simple correlation coefficients (e.g., Pearson or Spearman) are commonly used to identify the strength of empirical trends, they do not account for measurement uncertainties and/or model parameter covariance, and are therefore not appropriate for quantifying the primary correlations in this work.

Instead, we adopt the methods of \citet{2018ApJ...866...75W} and compare the
median fit slope with its 1$\sigma$ uncertainty (i.e., $|\beta|/\sigma_{\beta}$), which is the number of $\sigma$ by which the slope differs from zero. In cases in which the median fit slope has asymmetrical 1$\sigma$ uncertainties, we compare the median fit slope to the average of the asymmetrical 1$\sigma$ uncertainties. Using the levels of confidence defined in \citetalias{2018ApJ...866...75W}, we interpret 2–3$\sigma$ evidence values (i.e., $|\beta|/\sigma_{\beta}=[2,3]$) as marginal evidence and 3–5$\sigma$ evidence values ( i.e., $|\beta|/\sigma_{\beta}=[3,5]$) as evidence.

\subsection{Correlations between f and AGN/BLR parameters}
\label{sec:fvsparam}
Considering only correlations with at least marginal evidence, we find the following correlations with the virial coefficient, expressed as:

\begin{center}
    $\log_{10}(f)= \alpha + \beta\ X +\mathcal{N}(0,\sigma_{\rm int}^2)$
\end{center}
where $X$ represents the parameter of interest. 

\begin{enumerate}
    \item \textbf{Black Hole Mass:}
    \begin{center}
        \logfrmssigma\ vs. $\log_{10}(M_{\rm{BH}}/M_{\odot})$ \\
        $\alpha=-1.14 \pm 0.90$, $\beta=0.24 \pm 0.12$, $\sigma_{\rm int}=0.23 \pm 0.06$, \\
        2.0$\sigma$ marginal evidence 
        \vspace{0.2cm} \\
        


    \end{center}

        


    \item \textbf{Opening Angle (BLR disk thickness):}
    \begin{center}
        \logfrmsfwhm\ vs. $\log_{10}(\theta_o/{\rm deg.})$ \\
        $\alpha=1.61^{+0.51}_{-0.55}$, $\beta=-1.01^{+0.37}_{-0.34}$, $\sigma_{\rm int}=0.18 \pm 0.04$ \\
        $2.8 \sigma$ marginal evidence \vspace{0.2cm} \\

          
        \logfmeanfwhm\ vs. $\log_{10}(\theta_o/{\rm deg.})$ \\
$\alpha=1.27^{+0.49}_{-0.55}$, $\beta=-0.91^{+0.36}_{-0.33}$, $\sigma_{\rm int}=0.18 \pm 0.04$ \\

         $2.6\sigma$ marginal evidence
    \end{center}

    \item \textbf{Inclination Angle:}
    \begin{center}
        \logfrmssigma\ vs. $\log_{10}(\theta_i/{\rm deg.})$ \\
        $\alpha=2.87 ^{+1.10}_{-0.96}$, $\beta=-1.55^{+0.66}_{-0.76}$, $\sigma_{\rm int}=0.17 \pm 0.04$ \\
        $2.2\sigma$ marginal evidence \vspace{0.2cm}\\
        
        \logfmeanfwhm\ vs. $\log_{10}(\theta_i/{\rm deg.})$ \\
        $\alpha=2.06^{+0.93}_{-0.83}$, $\beta=-1.51^{+0.58}_{-0.65}$, $\sigma_{\rm int}=0.15 \pm 0.04$ \\
        $2.5\sigma$ marginal evidence\\
    \end{center}
\end{enumerate}

Figures \ref{fig:f_correlations_rms} and \ref{fig:f_correlations_mean} show the results for the rms and mean spectra, respectively, and linear regression fitting parameters are found in Table \ref{tab:table_mean_correlations}. For visual clarity, the error bars reflect the 68\% confidence intervals of the model estimates, rather than the 2D posterior distributions for model parameters as done in prior work \citep[e.g.,][]{2018ApJ...866...75W, 2023ApJ...948...95V}. Versions of Figures \ref{fig:f_correlations_rms} and \ref{fig:f_correlations_mean} that include the 2D posteriors can be found in Figures \ref{fig:f_correlations_rms_potato} and \ref{fig:f_correlations_mean_potato},  in the Appendix. 

\begin{deluxetable*}{c l c cccccc}
\tablecaption{Linear regression results for virial coefficient vs. BLR/AGN parameters}
\tablewidth{0pt}
\setlength{\tabcolsep}{5pt}
\tablehead{
\colhead{$f$-type} &
\colhead{ } &
\colhead{$\log_{10}(\rm{M}_{\rm BH}/\rm{M}_{\odot})$ } &
\colhead{$\log_{10}(\rm{L}_{5100}/{\rm erg~s}^{-1})$ } &
\colhead{$\log_{10}(\rm{L}_{\rm bol}/\rm{L}_{\rm Edd})$ } &
\colhead{$\log_{10}(\theta_o/{\rm deg.})$} &
\colhead{$\log_{10}(\theta_i/{\rm deg.})$} 
}
\startdata
& $\alpha$        &     $-1.14 \pm 0.90$ &  $-1.48^{+3.12}_{-2.93}$ &   $0.65 \pm 0.14$ &   $1.73^{+0.62}_{-0.67}$ &   $2.87^{+1.10}_{-0.96}$ \\
${\rm rms},{\sigma}$ & $\beta$           &   $0.24 \pm 0.12$ &          $0.05 \pm 0.07$ &  $-0.02 \pm 0.10$ &  $-0.72^{+0.44}_{-0.40}$ &  $-1.55^{+0.66}_{-0.76}$ \\
 & $\sigma_{\rm int}$ &   $0.23 \pm 0.05$ &          $0.23 \pm 0.06$ &   $0.24 \pm 0.06$ &          $0.22 \pm 0.05$ &          $0.17 \pm 0.04$ \\
 \hline
 \multirow{ 3}{*}{\begin{tabular}{l}rms, \\ FWHM\end{tabular}}
 & $\alpha$           &  $-0.09^{+1.01}_{-0.98}$ &  $-2.48^{+3.02}_{-2.90}$ &  $0.26 \pm 0.14$ &   $1.61^{+0.51}_{-0.55}$ &   $1.84^{+1.14}_{-1.44}$ \\
 & $\beta$              &                 $0.03 \pm 0.14$ &          $0.06 \pm 0.07$ &  $0.11 \pm 0.10$ &  $-1.01^{+0.37}_{-0.34}$ &  $-1.22^{+1.00}_{-0.79}$ \\
&$\sigma_{\rm int}$ &       $0.23 \pm 0.06$ &          $0.22 \pm 0.07$ &  $0.22 \pm 0.08$ &          $0.18 \pm 0.04$ &          $0.17 \pm 0.04$ \\
\hline
 \multirow{ 3}{*}{\begin{tabular}{l}mean, \\ $\sigma$ \end{tabular}}  
 & $\alpha$              &  $-1.05 \pm 0.95$ &  $0.03^{+3.20}_{-3.13}$ &   $0.49 \pm 0.13$ &   $1.73^{+0.59}_{-0.66}$ &   $2.43 \pm 1.21$ \\
&$\beta$               &   $0.22 \pm 0.13$ &         $0.01 \pm 0.07$ &  $-0.06 \pm 0.10$ &  $-0.79^{+0.43}_{-0.39}$ &  $-1.32 \pm 0.84$ \\
&$\sigma_{\rm {int}}$ &    $0.26 \pm 0.05$ &         $0.26 \pm 0.06$ &   $0.26 \pm 0.05$ &          $0.23 \pm 0.04$ &   $0.20 \pm 0.04$ \\ \hline
\multirow{ 3}{*}{\begin{tabular}{l}mean, \\ FWHM \end{tabular}}  &  $\alpha$            &  $-0.88^{+0.93}_{-0.91}$ &  $-3.42^{+2.90}_{-2.74}$ &         $0.03 \pm 0.14$ &   $1.27^{+0.49}_{-0.55}$ &   $2.06^{+0.93}_{-0.83}$ \\
 & $\beta$             &        $0.10 \pm 0.13$ &          $0.08 \pm 0.07$ &         $0.10 \pm 0.10$ &  $-0.91^{+0.36}_{-0.33}$ &  $-1.51^{+0.58}_{-0.65}$ \\
& $\sigma_{\rm {int}}$ &    $0.23 \pm 0.05$ &          $0.21 \pm 0.06$ &  $0.21^{+0.07}_{-0.06}$ &          $0.18 \pm 0.04$ &          $0.15 \pm 0.04$ \\
\hline \hline
\enddata
\tablecomments{Linear regression results used to determine correlations between the virial coefficient $f$ and select BLR/AGN shown in Figures \ref{fig:f_correlations_rms} and \ref{fig:f_correlations_mean}. The parameters $\alpha$ and $\beta$ represent the constant and slope of the linear regression, respectively. While $\sigma_{\rm{int}}$ represents the standard deviation of the intrinsic scatter. The corresponding relationship is therefore given by $\log_{10}f= \alpha + \beta\times \rm{parameter} +\mathcal{N}(0,\sigma_{\rm int})^2$.}
\label{tab:table_mean_correlations}
\end{deluxetable*}

Altogether, our results show systematic variation in $f$ with physical properties such as \mbh\,and BLR geometry -- reinforcing the need for an observationally motivated proxy for estimating individual AGN-specific virial coefficients. In particular, inclination angle has been measured for a number of sources using radio jets \citep[e.g.,][]{2005AJ....130.1418J, Agudo_2012}, but such measurements are not possible for all cases. Additionally, \citet{2026ApJ..1000...48B} recently showed that host galaxy inclination does not correlate with BLR inclination angle in their sample of Seyfert galaxies, suggesting there is no simple observable proxy. This further supports the reasoning of \citetalias{2023ApJ...948...95V}, which claims the correlations found in this section (i.e. with $\theta_o$ and $\theta_i$) lack any real utility,  motivating the study of line profile shape as an observational proxy for $f$, based on previous work of \citet{2006A&A...456...75C}.

\begin{figure*}[ht]
    \centering
    \begin{minipage}[b]{1.0\textwidth}
        \centering
        \includegraphics[width=1.0\textwidth]{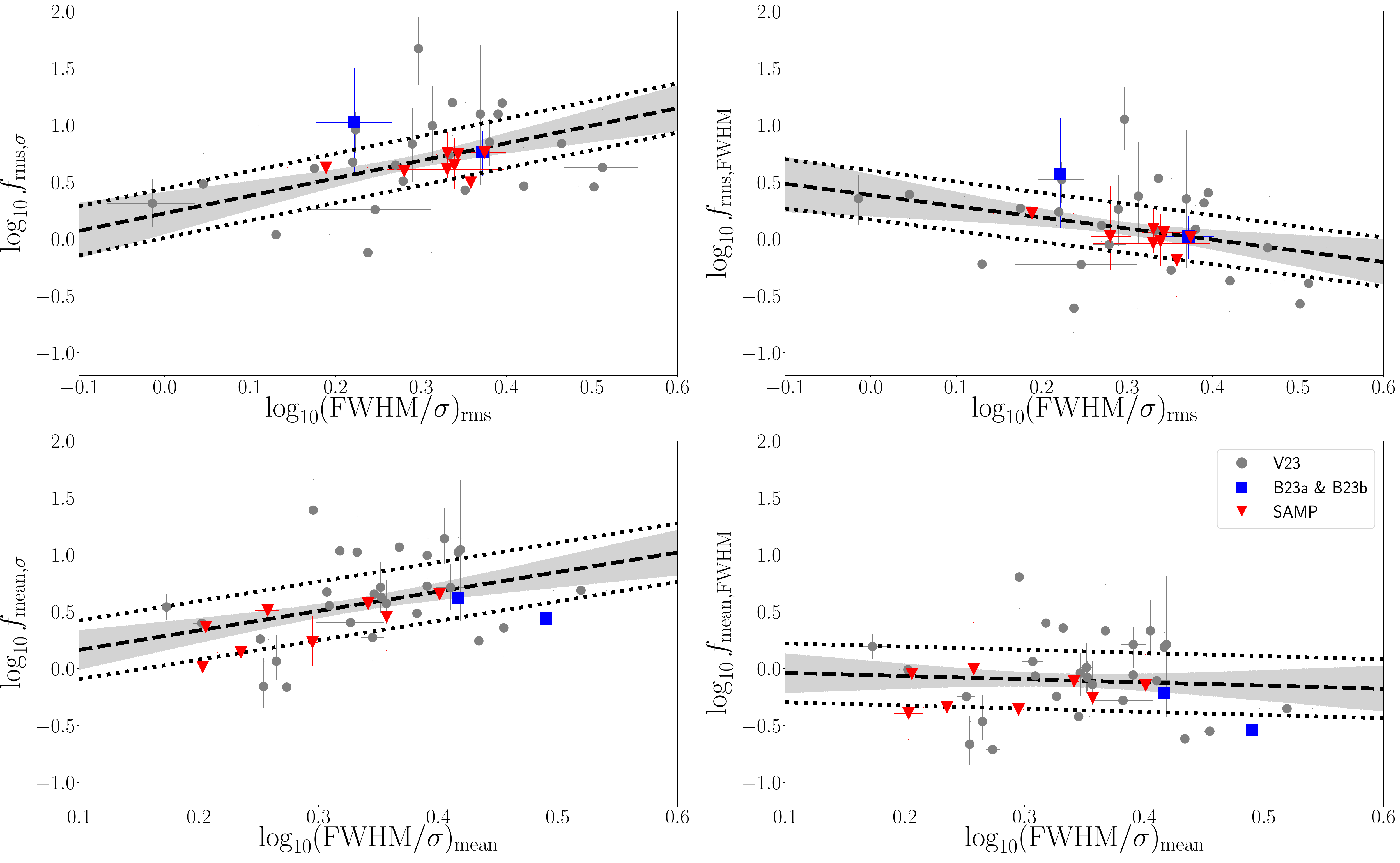}
        \caption{Correlations between rms (top) and mean (bottom) line-profile shape and virial coefficient. The dashed black lines and gray shaded regions give the median and 68\% confidence intervals of the linear regression. Dotted lines are offset above and below the dashed line by the median value of the intrinsic scatter. This work includes the addition of 10 new sources -- \citetalias{2023ApJ...959...25B} and \citetalias{2023ApJ...944...29B} are indicated by blue squares, and SAMP \citepalias{Wang_2026} are shown in red triangles.}
        \label{fig: mean_line_profile_shape_withf}
    \end{minipage}
      
    
    \begin{minipage}[b]{1.0\textwidth}
        \centering
        \includegraphics[width=\textwidth]{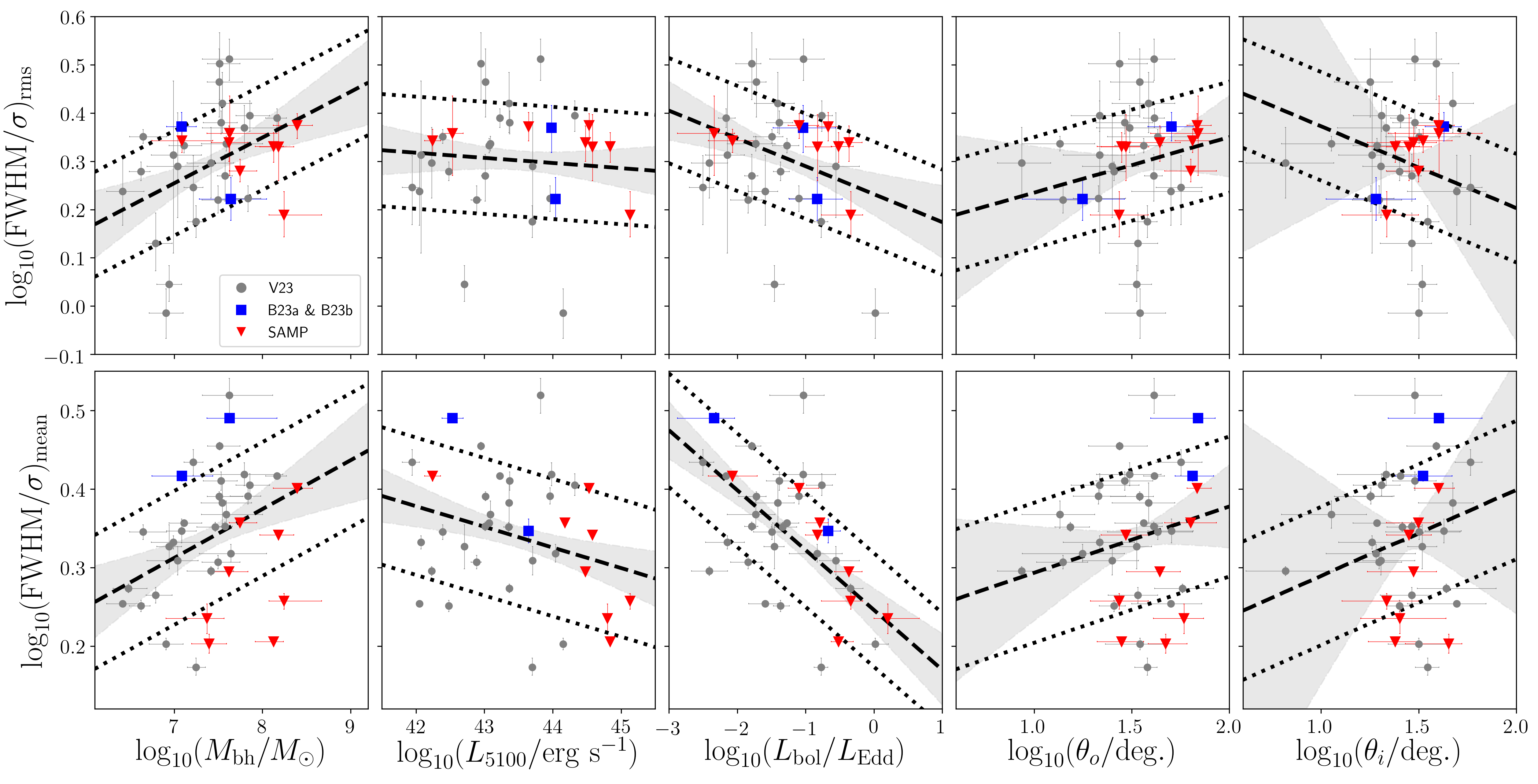}
        \caption{Correlations between line profile shape and \Hb\ BLR/AGN parameters using both the rms (top) and mean (bottom) spectrum. From left to right: \mbh, \Hb-emitting BLR inclination angle, \Hb-emitting BLR opening angle (disk thickness), Eddington ratio, and our ``inflow-outflow'' parameter. The dashed black lines and gray shaded regions give the median and 68\% confidence intervals of the linear regression. Dotted lines are offset above and below the dashed line by the median value of the intrinsic scatter. Grey points represent the  sample used in \citetalias{2023ApJ...948...95V}. This work includes the addition of 10 new sources -- \citetalias{2023ApJ...959...25B} and \citetalias{2023ApJ...944...29B} are indicated by blue squares, and SAMP \citepalias{Wang_2026} are shown in red triangles.}
        \label{fig:lineprofile-vs-params}
    \end{minipage}
\end{figure*}

\subsection{Correlations between $f$ and \lineprofile}
\label{sec: scalefactor_lineprofileshape}

In \citetalias{2023ApJ...948...95V}, we reported marginal evidence for a correlation between \logfrmssigma\ and \lineprofilerms, suggesting that line profile shape, quantified as $\log_{10}(\rm{FWHM}/\sigma)$, may encode BLR properties that influence the virial coefficient. This is promising because line profile shape is a measurable quantity, making it an appealing empirical proxy for $f$. With the inclusion of ten new sources, we test whether this trend persists in a larger sample. We fit the following relation:
\begin{center}
    $\log_{10}(f) = \alpha + \beta \log_{10}(\rm{FWHM}/\sigma) + \mathcal{N}(0, \sigma_{\rm int}^2)$
\end{center}
and find:
\begin{center}
    \textbf{\logfrmssigma\ vs. \lineprofilerms:} \\
    $\alpha=0.24 \pm 0.18$, $\beta=1.49^{+0.60}_{-0.62}$,
    $\sigma_{\rm int}=0.19 \pm 0.06$, \\
    $2.4\sigma$ marginal evidence \vspace{0.2cm}\\

    \textbf{\logfmeansigma\ vs. \lineprofilemean:} \\
    $\alpha=-0.02\pm 0.24$, $\beta=1.74\pm0.70$,
    $\sigma_{\rm int}=0.23 \pm 0.05$ \\
    $2.5\sigma$ marginal evidence
\end{center}

\begin{deluxetable}{c l c cccccc}
\tablecaption{Linear regression results for line profile shape \\ vs. virial coefficient}
\setlength{\tabcolsep}{6pt}
\tablehead{
\colhead{Line Profile Shape} &
\colhead{ } &
\colhead{$\log_{10}f_{\sigma}$} &
\colhead{$\log_{10}f_{\rm{FWHM}}$}
}
\startdata
\multirow{ 3}{*}{\begin{tabular}{l} $\log_{10}\Big(\frac{\rm{FWHM}}{\sigma}\Big)_{\rm{rms}}$ \\ \end{tabular}}  & $\alpha$           &      $0.24 \pm 0.18$ &      $0.39 \pm 0.19$ \\
 & $\beta$              & $1.49^{+0.60}_{-0.62}$ &     $-0.99 \pm 0.61$ \\
 & $\sigma_{\rm int}$  &        $0.19 \pm 0.06$ &      $0.19 \pm 0.06$ \\
\hline
\multirow{ 3}{*}{\begin{tabular}{l} $\log_{10}\Big(\frac{\rm{FWHM}}{\sigma}\Big)_{\rm{mean}}$ \\ \end{tabular}}  & 
$\alpha$ &       $-0.02 \pm 0.24$ &       $-0.001 \pm 0.244$ \\
 & $\beta$ &  $1.74 \pm 0.70$ &  $-0.31^{+0.73}_{-0.71}$ \\
 & $\sigma_{\rm int}$ & $0.23 \pm 0.05$ &          $0.23 \pm 0.05$ \\
 \hline \hline
\enddata
\tablecomments{Linear regression results for line profile shape vs. scale factor. The parameters $\alpha$ and $\beta$ represent the constant and slope of the regression, respectively, while $\sigma_{\rm{int}}$ represents the standard deviation of the intrinsic scatter. The corresponding relationship is therefore given by $\log_{10}(\textit{f})= \alpha + \beta  \log_{10}(\rm{FWHM/}\sigma) +\mathcal{N}(0,\sigma_{\rm int}^2)$.}
\label{tab:mean_lineprofile_f_linear_regression}
\end{deluxetable}

Linear regression results are found in Figure \ref{fig: mean_line_profile_shape_withf} and Table \ref{tab:mean_lineprofile_f_linear_regression}. These results reinforce the tentative correlation between the virial coefficient and line profile shape, $\log_{10}(\mathrm{FWHM}/\sigma)$ found in previous work. While the statistical significance remains marginal, the trend is recovered in an expanded sample that includes ten new AGN with an increased range in \mbh. This consistent result supports the idea that the shape of the broad H$\beta$ line may encode information about the geometry and kinematics of the BLR that directly influences the virial coefficient used in \mbh\ estimators. 

It is worth noting that the line profile shapes of the new sources largely fall within the same range covered by our previous sample, with $\log_{10}(\mathrm{FWHM}/\sigma) \sim [0.15 - 0.55]$. This range is centered near Gaussian-like profiles ($\log{10}(\mathrm{FWHM}/\sigma) \approx 0.37$), and does not significantly expand the dynamic range toward more extreme values. To fully characterize the slope and scatter of this relation, future BLR modeling of AGN with lower $\log_{10}(\mathrm{FWHM}/\sigma)$ is needed. 

\subsection{What drives line profile shape?}
\label{sec:lineprofile-vs-param}
\begin{deluxetable*}{c l c cccccc}
\tablecaption{Linear regression results for line profile shape vs. BLR/AGN parameters}
\tablewidth{0pt}
\setlength{\tabcolsep}{8pt}
\tablehead{
\colhead{Line Profile Shape} &
\colhead{ } &
\colhead{$\log_{10}(M_{\rm bh}/M_{\odot})$ } &
\colhead{$\log_{10}(\mathrm{L}_{5100}/{\rm erg~s}^{-1})$ } &
\colhead{$\log_{10}(L_{\rm bol}/L_{\rm Edd})$} &
\colhead{$\log_{10}(\theta_o/{\rm deg.})$} &
\colhead{$\log_{10}(\theta_i/{\rm deg.})$}
}
\startdata
\multirow{3}{*}{\begin{tabular}{l} $\log_{10}\Big(\frac{\rm{FWHM}}{\sigma}\Big)_{\rm{rms}}$ \end{tabular}}
 & $\alpha$            & $-0.40 \pm 0.37$ & $0.76^{+1.02}_{-1.01}$ & $0.23 \pm 0.04$ & $0.12^{+0.27}_{-0.29}$ & $0.54^{+0.79}_{-0.57}$ \\
 & $\beta$             & $0.09 \pm 0.05$  & $-0.01 \pm 0.02$       & $-0.06 \pm 0.03$ & $0.12 \pm 0.18$ & $-0.17^{+0.40}_{-0.54}$ \\
 & $\sigma_{\rm int}$  & $0.10 \pm 0.01$  & $0.11 \pm 0.02$        & $0.10 \pm 0.01$ & $0.11 \pm 0.02$ & $0.10 \pm 0.02$ \\
\hline
\multirow{3}{*}{\begin{tabular}{l} $\log_{10}\Big(\frac{\rm{FWHM}}{\sigma}\Big)_{\rm{mean}}$ \end{tabular}}
 & $\alpha$            & $-0.12 \pm 0.25$ & $1.48 \pm 0.69$        & $0.25 \pm 0.03$ & $0.21 \pm 0.17$ & $0.20^{+0.44}_{-0.46}$ \\
 & $\beta$             & $0.06 \pm 0.03$  & $-0.03 \pm 0.02$       & $-0.08 \pm 0.02$ & $0.08 \pm 0.11$ & $0.09^{+0.31}_{-0.30}$ \\
 & $\sigma_{\rm int}$  & $0.08 \pm 0.01$  & $0.08 \pm 0.01$        & $0.07 \pm 0.01$ & $0.08 \pm 0.01$ & $0.08 \pm 0.01$ \\
\hline\hline
\enddata
\tablecomments{Linear regression results for line profile shape vs. BLR/AGN parameters using both the mean and rms spectrum. The parameter $\alpha$ represents the constant in the regression and $\beta$ represents the slope of the regression, while $\sigma_{\rm{int}}^2$ represents the variance of the intrinsic scatter. The corresponding relationship is therefore given by $\log_{10}(\rm{FWHM/}\sigma)= \alpha + \beta\times \rm{parameter}+\mathcal{N}(0,\sigma_{\rm int}^2)$.}
\label{tab:lineprofile_params_linear_regression}
\end{deluxetable*}
Previous \citetalias{2023ApJ...948...95V} study explored possible physical drivers using toy \texttt{CARAMEL} models and found that a combination of features such as small BLR radius, increased disk thickness, inflow/outflow motions, and increased turbulence could produce narrower line profiles. In a similar attempt to understand what drives the observed variation in line profile shape, we investigate correlations between $\log_{10}(\mathrm{FWHM}/\sigma)$ and AGN/BLR parameters (see Figure \ref{fig:lineprofile-vs-params} and Table \ref{tab:lineprofile_params_linear_regression} for linear regression results). Considering only correlations with at least marginal evidence, we find the following correlations with line-profile shape, expressed as:

\begin{center}
    $\log_{10}(\rm{FWHM}/\sigma) = \alpha + \beta\ X +\mathcal{N}(0,\sigma_{\rm int}^2)$
\end{center}
where $X$ represents the parameter of interest. 

\begin{enumerate}
    \item \textbf{Eddington Ratio:}
    \begin{center}
     \subitem \lineprofilemean\ vs. $\log_{10}(L_{\rm{bol}}/L_{\rm{Edd}})$ \\
    $\alpha=0.25\pm0.03$, $\beta=-0.08 \pm 0.02$, $\sigma_{\rm{int}}=0.07\pm 0.01$\\
    $4.0\sigma$ evidence \\
    \end{center}
\end{enumerate}
We find evidence for a tight anticorrelation ($\sigma_{\rm{int}} \leq 0.08$ dex) between line profile shape and Eddington ratio, with stronger evidence from previous findings in \citet{Villafana22} which found $2\sigma$ marginal evidence.

\begin{figure}[h!]
\centering
\includegraphics[height=8cm, keepaspectratio]{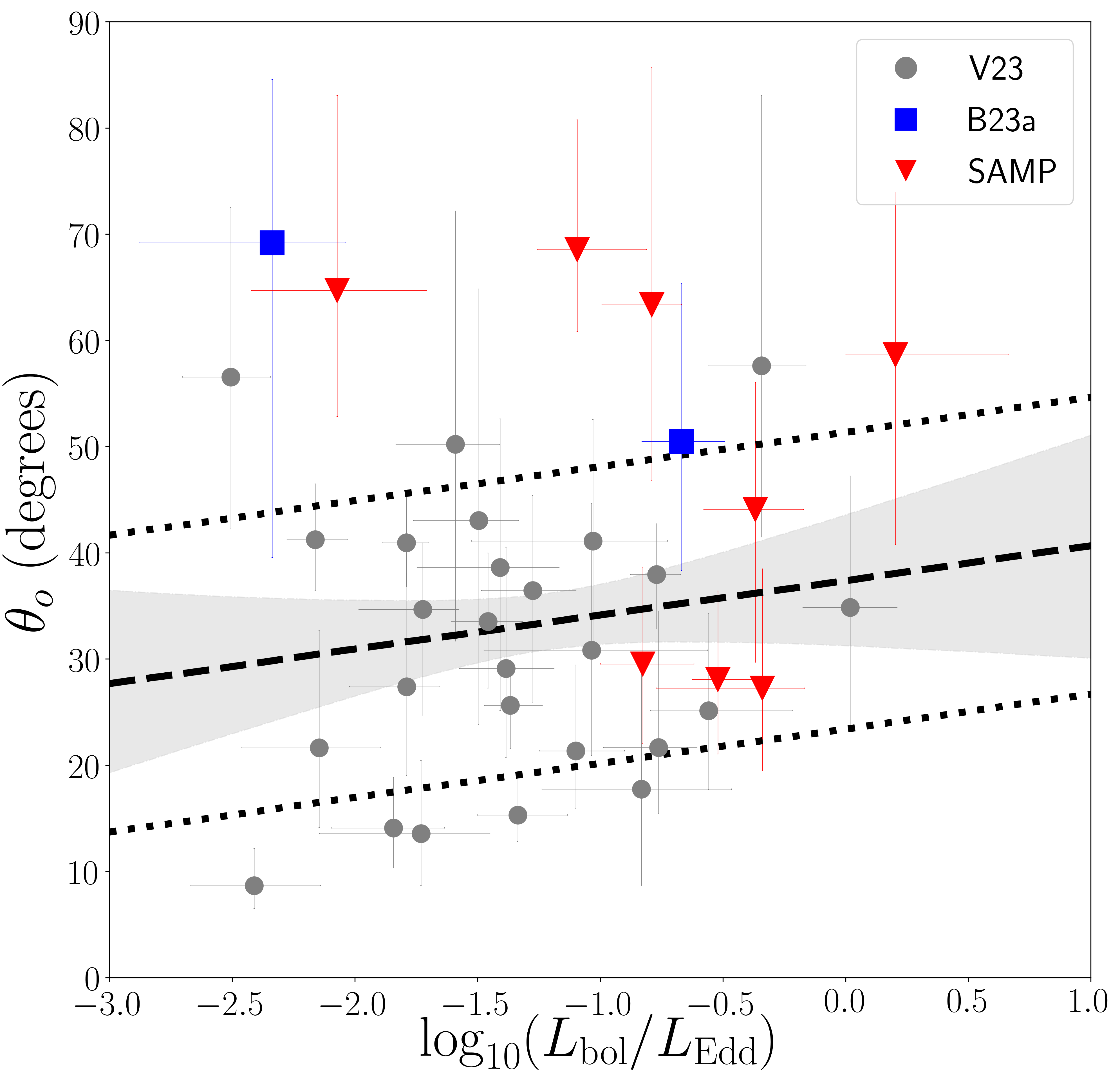}
\caption{Correlation between \Hb\ BLR opening angle (disk thickness) and Eddington ratio. The dashed black lines and gray shaded regions give the median and 68\% confidence intervals of the linear regression. Dotted lines are offset above and below the dashed line by the median value of the intrinsic scatter. Grey points represent the  sample used in \citetalias{2023ApJ...948...95V}. This work includes the addition of 10 new sources -- \citetalias{2023ApJ...959...25B} and \citetalias{2023ApJ...944...29B} are indicated by blue squares, and SAMP \citepalias{Wang_2026} are shown in red triangles}
\label{fig: thetaovseddington}
\end{figure}

\subsection{Partial Correlation Test}
Several of the AGN and BLR parameters considered in this work are not independent. Since $L_{\rm{bol}}/L_{\rm{Edd}}$ depends on both $L_{\rm{bol}} \approx 9 \lambda L_{\lambda}(5100 \mathrm{\AA})$  and \mbh, and \mbh\ and luminosity are known to be correlated \citep[e.g.][]{peterson04}, a correlation with Eddington ratio could in principle reflect contributions from either its luminosity or \mbh\ component, or a correlation with \mbh\ could be driven by luminosity. In particular, since $L_{\rm{bol}}/L_{\rm{Edd}}$ contains \mbh\ in the denominator, the anticorrelation detected between \lineprofilemean\ and $L_{\rm{bol}}/L_{\rm{Edd}}$ could in principle reflect the \mbh\ or luminosity dependence of Eddington ratio rather than accretion rate. Or alternatively, given the known correlation between black hole mass and luminosity, a trend between \mbh\ and $f$ could in principle reflect an underlying luminosity dependence.

To disentangle the interdependence of these parameters, we employ a partial correlation test using \textsc{pingouin} \citep{Vallat2018} to assess whether the trends reported in Section \ref{sec:RESULTS} may actually reflect the underlying interconnected nature of these parameters. We first compute correlation coefficients between each parameter with both \logfrmssigma\ and \lineprofilemean\ to establish a baseline before examining whether those trends persist after controlling for related parameters. We emphasize that the reported correlation coefficients and $p-$values serve here only as a diagnostic, as these tests do not account for correlated measurement uncertainties the way \texttt{linmix\_err} does, and therefore should not be used to quantify the strength of the observed trends.

For trends with \logfrmssigma, we find a correlation with \mbh\ ($r=0.450$, $p=0.0058$), but no correlation with $\lambda L_{\lambda}(5100 \mathrm{\AA})$ ($r=0.076, p=0.6577)$ or $L_{\rm{bol}}/L_{\rm{Edd}}$ ($r=-0.216, p=0.2048)$. Only \mbh\ shows a significant correlation with \logfrmssigma, and this trend persists after controlling for either luminosity ($r=0.503$, $p=0.0021$) or Eddington ratio ($r=0.480$, $p=0.0035$).  We note, however, that the $f$-\mbh\ correlation is expected since the two parameters are closely related; a larger $f$ should imply a larger \mbh\ by definition (Eqn. \ref{eq: rev_map}). Therefore, we conclude that the trend observed is independent of luminosity, but not informative as a physical trend.

Consistent with our findings in Section \ref{sec:lineprofile-vs-param}, we recover the anticorrelation with line profile shape and the Eddington ratio ($r=-0.537$ $p=0.0008$), but do not find a correlation with \mbh\ ($r=0.382, p=0.0213)$ or $\lambda L_{\lambda}(5100 \mathrm{\AA})$ ($r=-0.243, p=0.1538)$. Furthermore, the anticorrelation between line profile shape and Eddington ratio, not only survives after controlling for luminosity ($r=-0.627$, $p=0.0001$) and black hole mass ($r=-0.614$, $p=0.0001$), but strengthens, suggesting the trend is not induced by the dependence of Eddington ratio on either quantity.

Combined, these results are consistent with Eddington ratio being a primary driver of BLR geometry, with accretion rate influencing $f$ indirectly through that geometry rather than directly.

\section{Discussion}
\label{sec: discussion}
\subsection{Interpretation of trends}
The marginal correlation we recover between \lineprofile\ and the virial coefficient using a larger sample of AGN with BLR dynamical modeling, aligns with theoretical expectations that the shape of the broad emission line encodes BLR structural information. Importantly, this trend is recovered using individual AGN-specific virial coefficients inferred from dynamical modeling. This demonstrates that the relationship between line-profile shape and $f$ is not an artifact of the virial equation itself, but instead reflects underlying physical properties of the BLR, such as geometry, orientation, and/or accretion state. 

Interestingly, however, we do not find any significant correlation between line profile shape and inclination angle, suggesting that inclination angle is not the dominant driver of the observed
$f$–line profile shape relation. 
We find a strong anticorrelation between line profile shape and Eddington ratio, consistent with both our earlier findings \citep{Villafana22} and the intepretation of \citet{2006A&A...456...75C}, as well as recent FRADO simulations, which predict that the line profile shape decreases with increasing accretion rate \citep{Naddaf_2025}. However, we do not find any clear trends between $f$ and Eddington ratio, implying that accretion physics may only shape the virial coefficient indirectly. 

In particular, increased accretion rates may thicken the BLR vertically due to enhanced radiation pressure or increased turbulent support, leading to narrower, more centrally peaked line profiles. To further investigate this idea, we examined whether Eddington ratio correlates with BLR disk thickness,  $\theta_o$. Despite an apparent correlation (see Figure \ref{fig: thetaovseddington}), there is no statistical evidence. Nonetheless, these trends suggest a BLR structure dependent on accretion rate, though a precise physical mechanism remains uncertain.

\subsection{Implications for \mbh}
If both accretion rate and geometry contribute to variations in line profile shape and the virial coefficient, such differences across the AGN population can introduce biases in \mbh\ estimates, if using a sample average $\langle f \rangle$. In particular, these trends may have implications for populations such as narrow-line Seyfert 1 galaxies (NLS1s), which are often associated with high Eddington ratios and narrow line profiles \citep{2000MNRAS.314L..17M}.

The relevance of Eddington ratio on \mbh\ estimation was recently supported by the  work of \citet{Dalla-Bonta20}, who reanalyzed H$\beta$-based RM datasets and constructed updated single-epoch mass estimators. They find that the single-epoch mass residuals correlate strongly with Eddington ratio, reinforcing the idea that accretion physics may influence the BLR structure, and by extension, the line profile shape. Although they do not explicitly examine correlations between $f$ and line profile shape, their Eddington correction, derived from mass residuals, implicitly captures part of the same variations observed in this work.  More recently, \citet{2026arXiv260307047W} developed a new method to derive single epoch \mbh\ measurements by incorporating Eddington ratio as a new parameter to the R-L relation. Although they conclude more theoretical work is needed to understand the physical origin of the fundamental plane discovered, their work also suggests Eddington ratio may play a role in improving single epoch \mbh\ measurements. While we argue that line profile shape may serve as an empirical proxy for individual AGNs, our results suggest that Eddington ratio may contribute to shaping the line profile, and could therefore underlie the \lineprofile-$f$ relation we find.

Consistent with our findings, \citet{2025A&A...696A..30S} also emphasizes that line profile shape is an important parameter for \mbh\ estimation. Their spectroscopic survey of $z \sim 2$ AGN for GRAVITY+ follow-up revealed a subset of sources with strong non-Gaussian Balmer line profiles, typically exhibiting narrow cores and extended wings. They interpret these features as signatures of either a two-component BLR or a vertically extended, turbulence-dominated disk. They argue that when the profile shape of a target AGN differs significantly from that of the calibration sample, the resulting single-epoch mass estimate may be systematically biased. 
\subsection{Significance of tentative $\log_{10}(\mathrm{FWHM}/\sigma)$-$f$ trend}
Although the correlation between $\log_{10}(\mathrm{FWHM}/\sigma)$ and $f$ we have found remains marginal in significance, it is physically motivated. If confirmed with future BLR modeling across a wider range in line profile shapes \textbf{(e.g., NLS1s)}, this relation may offer an 
observationally driven method to improve \mbh\ estimates. This is particularly important for studies of high-redshift quasars during the era of JWST, where single-epoch \mbh\ estimates are often the only viable method for estimating \mbh, and several studies have reported that early SMBHs appear overmassive relative to their host galaxies \citep[e.g.][]{2022ApJ...941..106F, 2023ApJ...955L..24G, 2023ApJ...959...39H, 2023ApJ...957L...3P}. 

\section{Conclusions}
\label{sec: summary}
We present an updated compilation of 38 AGNs with \texttt{CARAMEL} H$\beta$ BLR dynamical modeling, the largest homogeneous sample to date, to investigate trends with derived AGN-specific virial coefficients and their implications for \mbh\ estimators. The main results are as follows.
\begin{enumerate}[(i)]
    \item We find marginal evidence for a correlation between the virial coefficient, $f$, and line profile shape, consistent with previous results, reinforcing the interpretation that line profile shape may encode physical BLR structural information that influence $f$.
    \item We recover previously reported trends between $f$ and BLR geometry, including correlations with opening and inclination angles. Interestingly, we do not find a significant correlation between inclination angle and line profile shape, suggesting inclination is not the dominant driver of the observed $f$–line profile shape relation.
    \item We find a tight anticorrelation between line profile shape and Eddington ratio, consistent with previous work, suggesting a  connection between accretion properties, line profile shape, and the inferred virial factor.
\end{enumerate}
Our results add to growing evidence \citep[e.g.,][]{2023ApJ...948...95V,2025A&A...696A..30S} that variations in line profile shape can introduce systematic biases in single-epoch \mbh\ estimates. Importantly, since line profile shape can be measured from single-epoch spectra, it offers a promising path for empirical correction of virial coefficients in large surveys and high-redshift studies. Future BLR dynamical modeling of reverberation mapping datasets targeting AGN with narrower \Hb\ profiles will be essential to test the robustness and broader applicability of a potential $\log_{10}(\rm{FWHM}/\sigma)$–$f$ relation. If confirmed, this relation may offer a viable empirical correction to refine \mbh\ estimates across cosmic time. 

\begin{acknowledgements}
We thank the anonymous referee for their helpful comments and suggestions, which significantly improved this manuscript. We also thank Thomas Lumley from The University of Auckland for the helpful discussion of linear regression fitting methods. MCB gratefully acknowledges support from the NSF through grant AST-2407802. JHW acknowledges the support by the Basic Science Research Program through the National Research Foundation of Korean Government (2021R1A2C3008486). VNB acknowledges funding support from STScI grant HST-GO-17103  and HST-AR-17063. \\

This research has made use of the NASA/IPAC Extragalactic Database (NED), which is operated by the Jet Propulsion Laboratory, California Institute of Technology, under contract with the National Aeronautics and Space Administration. 
\end{acknowledgements}

\begin{appendix}
\label{sec: appendix}

\begin{deluxetable*}{lcccccccc}
\tablecaption{Inferred Black Hole Mass and Virial Coefficient}
\tablewidth{0pt}
\setlength{\tabcolsep}{10pt}
\tablehead{
&\colhead{} &
\multicolumn{2}{c}{Rms } &
\multicolumn{2}{c}{Mean} \\
\colhead{Galaxy} & \colhead{$\log_{10}(M_{\rm bh}/M_{\odot})$} &  \colhead{$\log_{10}(f_{ {\rm rms},\sigma})$}  & \colhead{$\log_{10}(f_{ {\rm rms},{\rm FWHM}})$} & \colhead{$\log_{10}(f_{ {\rm mean},\sigma})$} & \colhead{$\log_{10}(f_{ {\rm mean},{\rm FWHM} })$}}
\startdata
$\rm 3C~120$ &  $7.84^{+0.15}_{-0.20}$ &   $0.96^{+0.15}_{-0.20}$ &     $0.52^{+0.16}_{-0.20}$ &   $0.99^{+0.15}_{-0.20}$ &   $0.21^{+0.15}_{-0.20}$ \\
           $\rm Arp~151$ &  $6.62^{+0.10}_{-0.13}$ &   $0.51^{+0.13}_{-0.16}$ &    $-0.05^{+0.13}_{-0.16}$ &   $0.26^{+0.13}_{-0.16}$ &  $-0.24^{+0.13}_{-0.16}$ \\
          $\rm IC~4329A$ &  $7.63^{+0.54}_{-0.26}$ &   $0.49^{+0.55}_{-0.26}$ &    $-0.19^{+0.57}_{-0.32}$ &   $0.44^{+0.55}_{-0.26}$ &  $-0.54^{+0.54}_{-0.26}$ \\
         $\rm J0140+234$ &  $8.25^{+0.43}_{-0.17}$ &   $0.63^{+0.40}_{-0.22}$ &     $0.22^{+0.41}_{-0.19}$ &   $0.51^{+0.41}_{-0.19}$ &  $-0.01^{+0.41}_{-0.18}$ \\
         $\rm J1026+523$ &  $7.62^{+0.22}_{-0.20}$ &   $0.65^{+0.24}_{-0.21}$ &    $-0.03^{+0.24}_{-0.20}$ &   $0.23^{+0.24}_{-0.20}$ &  $-0.37^{+0.24}_{-0.20}$ \\
         $\rm J1120+423$ &  $8.40^{+0.17}_{-0.27}$ &   $0.75^{+0.26}_{-0.30}$ &  $0.004^{+0.267}_{-0.294}$ &   $0.65^{+0.26}_{-0.30}$ &  $-0.16^{+0.26}_{-0.30}$ \\
         $\rm J1217+333$ &  $7.75^{+0.19}_{-0.12}$ &   $0.59^{+0.46}_{-0.29}$ &     $0.02^{+0.44}_{-0.29}$ &   $0.45^{+0.45}_{-0.29}$ &  $-0.26^{+0.46}_{-0.29}$ \\
         $\rm J1540+355$ &         $7.40 \pm 0.20$ &   $0.30^{+0.28}_{-0.23}$ &    $-0.26^{+0.27}_{-0.22}$ &   $0.01^{+0.27}_{-0.22}$ &  $-0.40^{+0.28}_{-0.22}$ \\
      $\rm MCG+04-22-04$ &  $7.59^{+0.42}_{-0.28}$ &   $1.21^{+0.40}_{-0.31}$ &     $0.54^{+0.40}_{-0.31}$ &   $1.08^{+0.40}_{-0.31}$ &   $0.34^{+0.40}_{-0.31}$ \\
        $\rm Mrk ~ 1048$ &  $7.79^{+0.44}_{-0.48}$ &   $1.05^{+0.63}_{-0.57}$ &     $0.33^{+0.64}_{-0.59}$ &   $1.01^{+0.62}_{-0.58}$ &   $0.16^{+0.63}_{-0.57}$ \\
        $\rm Mrk ~ 1392$ &  $8.16^{+0.11}_{-0.13}$ &   $1.09^{+0.13}_{-0.14}$ &     $0.31^{+0.13}_{-0.14}$ &          $1.02 \pm 0.13$ &          $0.18 \pm 0.14$ \\
         $\rm Mrk ~ 841$ &  $7.62^{+0.50}_{-0.30}$ &   $0.64^{+0.49}_{-0.40}$ &    $-0.38^{+0.48}_{-0.41}$ &   $0.70^{+0.48}_{-0.39}$ &  $-0.35^{+0.49}_{-0.39}$ \\
          $\rm Mrk~1310$ &  $7.42^{+0.26}_{-0.27}$ &          $1.65 \pm 0.29$ &            $1.05 \pm 0.28$ &   $1.40^{+0.26}_{-0.29}$ &   $0.80^{+0.26}_{-0.28}$ \\
           $\rm Mrk~141$ &  $7.46^{+0.15}_{-0.21}$ &          $0.71 \pm 0.23$ &            $0.71 \pm 0.23$ &          $0.70 \pm 0.23$ &        $0.004 \pm 0.222$ \\
          $\rm Mrk~1501$ &  $7.86^{+0.22}_{-0.16}$ &   $1.19^{+0.27}_{-0.21}$ &     $0.41^{+0.26}_{-0.22}$ &   $1.14^{+0.27}_{-0.21}$ &   $0.32^{+0.27}_{-0.21}$ \\
          $\rm Mrk~1511$ &  $7.11^{+0.20}_{-0.17}$ &   $0.74^{+0.21}_{-0.19}$ &     $0.08^{+0.20}_{-0.19}$ &   $0.58^{+0.20}_{-0.19}$ &         $-0.14 \pm 0.20$ \\
           $\rm Mrk~279$ &         $7.58 \pm 0.08$ &   $0.65^{+0.14}_{-0.12}$ &     $0.11^{+0.17}_{-0.15}$ &   $0.63^{+0.14}_{-0.12}$ &  $-0.08^{+0.14}_{-0.12}$ \\
           $\rm Mrk~335$ &         $7.25 \pm 0.11$ &          $0.62 \pm 0.12$ &            $0.27 \pm 0.12$ &          $0.54 \pm 0.11$ &          $0.19 \pm 0.10$ \\
            $\rm Mrk~50$ &  $7.50^{+0.25}_{-0.18}$ &   $0.68^{+0.25}_{-0.22}$ &     $0.24^{+0.24}_{-0.21}$ &   $0.68^{+0.23}_{-0.21}$ &   $0.06^{+0.23}_{-0.21}$ \\
          $\rm NGC~3227$ &  $7.09^{+0.35}_{-0.34}$ &   $0.74^{+0.37}_{-0.35}$ &     $0.06^{+0.38}_{-0.35}$ &   $0.62^{+0.37}_{-0.35}$ &  $-0.21^{+0.37}_{-0.35}$ \\
          $\rm NGC~3783$ &  $7.51^{+0.26}_{-0.13}$ &   $0.84^{+0.26}_{-0.16}$ &  $-0.08^{+0.26}_{-0.20}$ &   $0.72^{+0.26}_{-0.14}$ &  $-0.06^{+0.26}_{-0.14}$ \\
          $\rm NGC~4151$ &         $7.21 \pm 0.11$ &   $0.25^{+0.13}_{-0.11}$ &    $-0.23^{+0.19}_{-0.17}$ &   $0.24^{+0.13}_{-0.11}$ &  $-0.62^{+0.13}_{-0.12}$ \\
          $\rm NGC~4593$ &  $6.65^{+0.27}_{-0.15}$ &   $0.42^{+0.28}_{-0.20}$ &    $-0.28^{+0.28}_{-0.20}$ &   $0.26^{+0.28}_{-0.20}$ &  $-0.43^{+0.28}_{-0.19}$ \\
          $\rm NGC~5548$ &  $7.51^{+0.23}_{-0.14}$ &   $0.47^{+0.31}_{-0.24}$ &    $-0.55^{+0.39}_{-0.23}$ &   $0.35^{+0.33}_{-0.20}$ &  $-0.56^{+0.32}_{-0.20}$ \\
  $\rm NGC~5548~(STORM)$ &  $7.54^{+0.34}_{-0.24}$ &   $0.47^{+0.33}_{-0.28}$ &    $-0.38^{+0.33}_{-0.25}$ &   $0.48^{+0.33}_{-0.24}$ &  $-0.28^{+0.32}_{-0.26}$ \\
          $\rm NGC~6814$ &  $6.42^{+0.24}_{-0.18}$ &  $-0.12^{+0.26}_{-0.22}$ &    $-0.60^{+0.26}_{-0.21}$ &  $-0.16^{+0.24}_{-0.19}$ &  $-0.66^{+0.24}_{-0.19}$ \\
     $\rm NPM1G+27.0587$ &  $7.64^{+0.40}_{-0.36}$ &   $0.99^{+0.48}_{-0.43}$ &     $0.53^{+0.49}_{-0.44}$ &   $1.02^{+0.47}_{-0.45}$ &   $0.39^{+0.47}_{-0.46}$ \\
     $\rm PG ~ 2209+184$ &  $7.53^{+0.19}_{-0.20}$ &          $0.84 \pm 0.21$ &     $0.08^{+0.21}_{-0.20}$ &   $0.71^{+0.21}_{-0.20}$ &         $-0.11 \pm 0.21$ \\
       $\rm PG~0947+396$ &  $8.18^{+0.17}_{-0.21}$ &   $0.61^{+0.26}_{-0.23}$ &    $-0.04^{+0.27}_{-0.25}$ &   $0.57^{+0.25}_{-0.23}$ &  $-0.11^{+0.25}_{-0.22}$ \\
       $\rm PG~1121+422$ &  $8.13^{+0.11}_{-0.21}$ &   $0.74^{+0.18}_{-0.21}$ &     $0.08^{+0.16}_{-0.21}$ &   $0.36^{+0.16}_{-0.21}$ &  $-0.05^{+0.16}_{-0.21}$ \\
       $\rm PG~1310-108$ &  $6.48^{+0.21}_{-0.18}$ &  $-0.16^{+0.31}_{-0.25}$ &    $-0.16^{+0.32}_{-0.25}$ &  $-0.16^{+0.32}_{-0.25}$ &  $-0.71^{+0.32}_{-0.25}$ \\
       $\rm PG~1427+480$ &  $7.37^{+0.19}_{-0.47}$ &          $0.29 \pm 0.44$ &            $0.01 \pm 0.41$ &          $0.11 \pm 0.41$ &  $-0.36^{+0.40}_{-0.42}$ \\
       $\rm PG~2130+099$ &         $6.91 \pm 0.19$ &          $0.31 \pm 0.21$ &            $0.35 \pm 0.21$ &   $0.39^{+0.21}_{-0.19}$ &         $-0.01 \pm 0.20$ \\
        $\rm RBS ~ 1303$ &  $6.79^{+0.19}_{-0.11}$ &   $0.03^{+0.27}_{-0.19}$ &    $-0.22^{+0.25}_{-0.18}$ &   $0.06^{+0.24}_{-0.17}$ &  $-0.46^{+0.24}_{-0.17}$ \\
        $\rm RBS ~ 1917$ &  $7.04^{+0.23}_{-0.35}$ &   $0.84^{+0.30}_{-0.35}$ &            $0.25 \pm 0.32$ &   $0.54^{+0.26}_{-0.32}$ &  $-0.07^{+0.25}_{-0.32}$ \\
 $\rm RXJ ~ 2044.0+2833$ &         $7.09 \pm 0.17$ &   $0.77^{+0.18}_{-0.20}$ &     $0.02^{+0.17}_{-0.20}$ &   $0.66^{+0.17}_{-0.20}$ &  $-0.04^{+0.17}_{-0.20}$ \\
          $\rm SBS~1116$ &  $6.99^{+0.32}_{-0.25}$ &   $0.99^{+0.36}_{-0.33}$ &     $0.38^{+0.46}_{-0.39}$ &   $1.03^{+0.30}_{-0.28}$ &   $0.37^{+0.30}_{-0.28}$ \\
            $\rm Zw~229$ &         $6.94 \pm 0.14$ &   $0.48^{+0.26}_{-0.22}$ &     $0.39^{+0.26}_{-0.22}$ &   $0.41^{+0.25}_{-0.22}$ &  $-0.24^{+0.26}_{-0.21}$ \\
\enddata
\tablecomments{Inferred \textsc{caramel} black hole masses and derived AGN-specific virial coefficients for the entire sample of sources with \textsc{caramel} BLR dynamical modeling. Column 1 lists the galaxy name, column 2 list the \textsc{caramel} \mbh\ estimate, as defined by the 68\% confidence interval of the resulting posterior distribution function. Columns 3, 4, 5, and 6, list the virial coefficient derived using the \textsc{caramel} \mbh\ estimate, the measured cross correlation time lag, and the line dispersion or FWHM line width measured using the rms or mean spectrum. }
\label{tab:table_individual_f}
\end{deluxetable*}
\begin{deluxetable*}{lccccccc}[h!]
\tablecaption{Line widths and line profile shapes for sources with \textsc{caramel} dynamical modeling}
\tablewidth{0pt}
\setlength{\tabcolsep}{7.5pt}
\tablehead{
\colhead{} &
\multicolumn{3}{c}{Rms } &
\multicolumn{3}{c}{Mean} &
\colhead{} \\ \hline
\colhead{Galaxy}  &\colhead{FWHM} & \colhead{$\sigma_{\rm{line}}$} & \colhead{$\log_{10}(\rm{FWHM}/\sigma)$} & \colhead{FWHM} & \colhead{$\sigma_{\rm {line}}$} & \colhead{$\log_{10}(\rm{FWHM}/\sigma)$} & \colhead{Ref.}
}
\startdata
            $\rm 3C~120$ &             $2035 \pm 97$ &               $1218 \pm 47$ &                          $0.22 \pm 0.03$ &              $2893 \pm 22$ &                $1175 \pm 26$ &                           $0.39 \pm 0.01$ &    b \\
           $\rm Arp~151$ &             $2458 \pm 82$ &               $1295 \pm 37$ &                          $0.28 \pm 0.02$ &              $3076 \pm 39$ &                $1726 \pm 17$ &                           $0.25 \pm 0.01$ &    a \\
          $\rm IC~4329A$ &            $4789 \pm 869$ &               $2112 \pm 93$ &                           $0.35\pm 0.08$ &              $6944 \pm 51$ &                 $2247 \pm 8$ &                          $0.49 \pm 0.003$ &    g \\
         $\rm J0140+234$ &             $2230 \pm 97$ &              $1438 \pm 138$ &                          $0.19 \pm 0.05$ &              $2896 \pm 62$ &                $1601 \pm 12$ &                           $0.26 \pm 0.01$ &    h \\
         $\rm J1026+523$ &            $2608 \pm 154$ &               $1198 \pm 66$ &                          $0.34 \pm 0.04$ &              $3822 \pm 31$ &                 $1938 \pm 7$ &                          $0.29 \pm 0.004$ &    h \\
         $\rm J1120+423$ &            $5158 \pm 278$ &               $2180 \pm 45$ &                          $0.37 \pm 0.02$ &              $6211 \pm 31$ &                $2466 \pm 11$ &                          $0.40 \pm 0.003$ &    h \\
         $\rm J1217+333$ &            $3233 \pm 154$ &               $1691 \pm 54$ &                          $0.28 \pm 0.02$ &              $4476 \pm 31$ &                $1971 \pm 18$ &                          $0.36 \pm 0.004$ &    h \\
         $\rm J1540+355$ &             $2041 \pm 93$ &               $1077 \pm 79$ &                          $0.28 \pm 0.04$ &              $2383 \pm 62$ &                $1493 \pm 14$ &                           $0.20 \pm 0.01$ &    h \\
      $\rm MCG+04-22-04$ &             $2120 \pm 39$ &                $977 \pm 29$ &                          $0.34 \pm 0.01$ &              $2658 \pm 57$ &                $1141 \pm 39$ &                           $0.37 \pm 0.02$ &    f \\
        $\rm Mrk ~ 1048$ &            $4042 \pm 406$ &               $1726 \pm 76$ &                          $0.37 \pm 0.05$ &              $4830 \pm 80$ &                $1840 \pm 58$ &                           $0.42 \pm 0.02$ &    f \\
          $\rm Mrk~1310$ &            $1823 \pm 157$ &               $921 \pm 135$ &                          $0.30 \pm 0.07$ &              $2425 \pm 19$ &                $1229 \pm 12$ &                          $0.30 \pm 0.005$ &    a \\
        $\rm Mrk ~ 1392$ &            $3690 \pm 138$ &               $1501 \pm 38$ &                          $0.39 \pm 0.02$ &              $4267 \pm 25$ &                $1635 \pm 13$ &                          $0.42 \pm 0.004$ &    f \\
           $\rm Mrk~141$ &                   \nodata &                     \nodata &                                  \nodata &              $5129 \pm 45$ &                $2280 \pm 21$ &                           $0.35 \pm 0.01$ &    c \\
          $\rm Mrk~1501$ &            $3476 \pm 214$ &               $1401 \pm 48$ &                          $0.39 \pm 0.03$ &              $3780 \pm 25$ &                $1486 \pm 48$ &                           $0.41 \pm 0.01$ &    b \\
          $\rm Mrk~1511$ &             $3236 \pm 65$ &               $1506 \pm 42$ &                          $0.33 \pm 0.01$ &              $4154 \pm 28$ &                $1828 \pm 12$ &                          $0.36 \pm 0.004$ &    c \\
           $\rm Mrk~279$ &            $3306 \pm 338$ &               $1778 \pm 71$ &                          $0.27 \pm 0.05$ &              $4099 \pm 43$ &                $1821 \pm 13$ &                           $0.35 \pm 0.01$ &    c \\
           $\rm Mrk~335$ &             $1853 \pm 79$ &               $1239 \pm 78$ &                          $0.17 \pm 0.03$ &               $2018 \pm 1$ &                $1354 \pm 34$ &                           $0.17 \pm 0.01$ &    b \\
            $\rm Mrk~50$ &            $3355 \pm 128$ &              $2020 \pm 103$ &                          $0.22 \pm 0.03$ &              $4101 \pm 56$ &                $2024 \pm 31$ &                           $0.31 \pm 0.01$ &    c \\
         $\rm Mrk ~ 841$ &            $7452 \pm 660$ &               $2278 \pm 96$ &                          $0.51 \pm 0.04$ &             $7073 \pm 311$ &                $2139 \pm 55$ &                           $0.52 \pm 0.02$ &    f \\
          $\rm NGC~3227$ &          $3710 \pm 186$ &           $1682 \pm 39$ &                          $0.34 \pm 0.02$ &          $5070 \pm 69$ &            $1943 \pm 13$ &                           $0.42 \pm 0.01$ &    g \\
           $\rm NGC~3783$ &            $4728 \pm 676$ &              $1619 \pm 137$ &                          $0.47 \pm 0.07$ &              $4486 \pm 35$ &                $1825 \pm 19$ &                           $0.39 \pm 0.01$ &    e \\
          $\rm NGC~4151$ &            $4711 \pm 750$ &               $2680 \pm 64$ &                          $0.25 \pm 0.07$ &             $7382 \pm 279$ &                $2724 \pm 17$ &                           $0.43 \pm 0.02$ &    g \\
          $\rm NGC~4593$ &             $3597 \pm 72$ &               $1601 \pm 40$ &                          $0.35 \pm 0.01$ &              $4264 \pm 41$ &                $1925 \pm 38$ &                           $0.35 \pm 0.01$ &    c \\
          $\rm NGC~5548$ &          $12539 \pm 1927$ &              $3900 \pm 266$ &                          $0.51 \pm 0.07$ &            $12402 \pm 111$ &                $4354 \pm 25$ &                          $0.45 \pm 0.005$ &    a \\
  $\rm NGC~5548~(STORM)$ &           $10861 \pm 739$ &              $4115 \pm 513$ &                          $0.42 \pm 0.07$ &             $9612 \pm 427$ &               $3983 \pm 150$ &                           $0.38 \pm 0.02$ &    d \\
          $\rm NGC~6814$ &            $2945 \pm 283$ &              $1697 \pm 224$ &                          $0.24 \pm 0.08$ &              $3129 \pm 14$ &                $1744 \pm 12$ &                          $0.25 \pm 0.004$ &    a \\
     $\rm NPM1G+27.0587$ &            $2893 \pm 177$ &              $1735 \pm 136$ &                          $0.22 \pm 0.04$ &              $3501 \pm 28$ &                $1683 \pm 42$ &                           $0.32 \pm 0.01$ &    f \\
       $\rm PG~0947+396$ &            $4910 \pm 648$ &              $2294 \pm 192$ &                          $0.33 \pm 0.06$ &              $5258 \pm 36$ &                $2396 \pm 16$ &                          $0.34 \pm 0.004$ &    h \\
       $\rm PG~1121+422$ &             $2230 \pm 62$ &               $1037 \pm 70$ &                          $0.33 \pm 0.03$ &              $2613 \pm 31$ &                $1627 \pm 10$ &                           $0.21 \pm 0.01$ &    h \\
       $\rm PG~1310-108$ &                   \nodata &                     \nodata &                                  \nodata &              $3422 \pm 21$ &                $1823 \pm 20$ &                           $0.27 \pm 0.01$ &    c \\
       $\rm PG~1427+480$ &             $1786 \pm 93$ &              $1292 \pm 155$ &                          $0.14 \pm 0.06$ &              $2721 \pm 97$ &                $1582 \pm 38$ &                           $0.24 \pm 0.02$ &    h \\
       $\rm PG~2130+099$ &            $1409 \pm 143$ &               $1459 \pm 93$ &                         $-0.01 \pm 0.05$ &              $2107 \pm 32$ &                $1321 \pm 11$ &                           $0.20 \pm 0.01$ &    b \\
     $\rm PG ~ 2209+184$ &             $3247 \pm 88$ &               $1353 \pm 64$ &                          $0.38 \pm 0.02$ &              $4045 \pm 34$ &                $1573 \pm 40$ &                           $0.41 \pm 0.01$ &    f \\
        $\rm RBS ~ 1303$ &            $1738 \pm 113$ &              $1292 \pm 156$ &                          $0.13 \pm 0.06$ &              $2286 \pm 21$ &                $1243 \pm 26$ &                           $0.26 \pm 0.01$ &    f \\
        $\rm RBS ~ 1917$ &            $1653 \pm 287$ &               $851 \pm 154$ &                          $0.28 \pm 0.12$ &              $2399 \pm 11$ &                $1180 \pm 50$ &                           $0.31 \pm 0.02$ &    f \\
 $\rm RXJ ~ 2044.0+2833$ &             $2047 \pm 72$ &                $870 \pm 50$ &                          $0.37 \pm 0.03$ &              $2196 \pm 31$ &                 $989 \pm 32$ &                           $0.35 \pm 0.02$ &    f \\
          $\rm SBS~1116$ &           $3202 \pm 1127$ &              $1550 \pm 310$ &                           $0.31\pm 0.19$ &              $3135 \pm 36$ &                $1460 \pm 23$ &                           $0.33 \pm 0.01$ &    a \\
            $\rm Zw~229$ &             $1789 \pm 93$ &              $1609 \pm 109$ &                          $0.04 \pm 0.04$ &             $3705 \pm 203$ &                $1747 \pm 56$ &                           $0.33 \pm 0.03$ &    c \\
            \bottomrule
\enddata
\tablecomments{Line widths and line profile shapes for the entire sample of sources with \textsc{caramel} BLR dynamical modeling. All line widths are given in km s$^{-1}$. Column 1 lists the galaxy name, columns 2 and 3 list the FWHM and line dispersion line widths measured using the rms spectrum. Column 4 lists the associated line profile shape, i.e. ratio of line widths. Note: a line profile shape of $\log_{10}(\rm{FWHM}/\sigma)=0.371$ corresponds to a Gaussian profile, while $\log_{10}(\rm{FWHM}/\sigma) < 0.371$ corresponds to a Lorentz profile and $\log_{10}(\rm{FWHM}/\sigma) > 0.371$ corresponds to a flat topped profile. Columns 5 and 6 list the FWHM and line dispersion line widths measured using the mean spectrum, and column 7 list the corresponding line profile shape. Column 8 indicates the references for the line widths measurements found in columns 2, 3, 5, and 6. The references are coded as follows: (a) \citet{2012ApJ...747...30P}, (b) \citetalias{2023ApJ...948...95V}, (c) \citet{2015ApJS..217...26B}, (d) \citet{2017ApJ...837..131P}, (e) \citet{2021ApJ...906...50B}, (f) \citet{u2021lick}, (g) \citet{2006ApJ...651..775B}, (h) \citet{Woo24}}
\label{tab: line_widths}
\end{deluxetable*}
\begin{deluxetable}{llcccc}
\setlength{\tabcolsep}{9pt}
\tablecaption{Inferred \Hb\ BLR Inclination \& Disk Thickness}
\tablehead{
\colhead{Galaxy} & 
\colhead{$\theta_i$} & 
\colhead{$\theta_o$}}
\startdata
           $\rm 3C~120$ &    $18^{+6}_{-3}$ &    $21^{+8}_{-5}$ \\
           $\rm Arp~151$ &    $25^{+3}_{-3}$ &    $26^{+4}_{-4}$ \\
          $\rm IC~4329A$ &  $40^{+27}_{-18}$ &  $69^{+15}_{-30}$ \\
         $\rm J0140+234$ &   $22^{+10}_{-9}$ &   $27^{+11}_{-8}$ \\
         $\rm J1026+523$ &   $30^{+9}_{-12}$ &  $44^{+12}_{-14}$ \\
         $\rm J1120+423$ &   $40^{+8}_{-10}$ &   $69^{+12}_{-8}$ \\
         $\rm J1217+333$ &   $32^{+6}_{-10}$ &  $63^{+22}_{-17}$ \\
         $\rm J1540+355$ &   $45^{+7}_{-18}$ &   $47^{+14}_{-9}$ \\
      $\rm MCG+04-22-04$ &    $11^{+6}_{-5}$ &    $14^{+7}_{-5}$ \\
        $\rm Mrk ~ 1048$ &    $22^{+9}_{-9}$ &  $31^{+14}_{-10}$ \\
        $\rm Mrk ~ 1392$ &    $26^{+3}_{-3}$ &    $41^{+5}_{-5}$ \\
         $\rm Mrk ~ 841$ &  $30^{+11}_{-15}$ &  $41^{+11}_{-11}$ \\
          $\rm Mrk~1310$ &     $7^{+5}_{-2}$ &     $9^{+4}_{-2}$ \\
           $\rm Mrk~141$ &    $26^{+6}_{-4}$ &    $15^{+4}_{-2}$ \\
          $\rm Mrk~1501$ &    $20^{+5}_{-6}$ &   $22^{+13}_{-6}$ \\
          $\rm Mrk~1511$ &    $19^{+6}_{-5}$ &   $36^{+9}_{-11}$ \\
           $\rm Mrk~279$ &        $29 \pm 3$ &    $41^{+4}_{-4}$ \\
           $\rm Mrk~335$ &    $35^{+5}_{-5}$ &    $38^{+5}_{-5}$ \\
            $\rm Mrk~50$ &    $20^{+6}_{-5}$ &    $14^{+5}_{-4}$ \\
          $\rm NGC~3227$ &   $33^{+14}_{-9}$ &  $65^{+18}_{-12}$ \\
          $\rm NGC~3783$ &    $18^{+5}_{-6}$ &   $35^{+6}_{-10}$ \\
          $\rm NGC~4151$ &   $58^{+8}_{-10}$ &  $57^{+16}_{-14}$ \\
          $\rm NGC~4593$ &  $32^{+20}_{-10}$ &  $43^{+22}_{-19}$ \\
          $\rm NGC~5548$ &  $39^{+12}_{-11}$ &   $27^{+11}_{-8}$ \\
  $\rm NGC~5548~(STORM)$ &  $47^{+13}_{-16}$ &  $39^{+14}_{-13}$ \\
          $\rm NGC~6814$ &  $49^{+20}_{-22}$ &  $50^{+22}_{-19}$ \\
     $\rm NPM1G+27.0587$ &   $19^{+11}_{-8}$ &   $18^{+11}_{-9}$ \\
     $\rm PG ~ 2209+184$ &    $30^{+9}_{-7}$ &   $29^{+11}_{-8}$ \\
       $\rm PG~0947+396$ &    $28^{+9}_{-7}$ &    $30^{+9}_{-7}$ \\
       $\rm PG~1121+422$ &    $24^{+8}_{-7}$ &    $28^{+8}_{-7}$ \\
       $\rm PG~1310-108$ &  $44^{+35}_{-13}$ &  $58^{+25}_{-16}$ \\
       $\rm PG~1427+480$ &   $25^{+18}_{-9}$ &  $59^{+15}_{-18}$ \\
       $\rm PG~2130+099$ &  $32^{+13}_{-10}$ &  $35^{+12}_{-12}$ \\
        $\rm RBS ~ 1303$ &    $29^{+8}_{-9}$ &   $34^{+9}_{-10}$ \\
        $\rm RBS ~ 1917$ &   $20^{+10}_{-4}$ &    $25^{+9}_{-7}$ \\
 $\rm RXJ ~ 2044.0+2833$ &   $43^{+10}_{-8}$ &  $50^{+15}_{-12}$ \\
          $\rm SBS~1116$ &    $18^{+8}_{-6}$ &   $22^{+11}_{-8}$ \\
            $\rm Zw~229$ &    $33^{+6}_{-5}$ &    $34^{+6}_{-6}$ \\   \hline 
\enddata
\tablecomments{Full sample of sources with \textsc{caramel} BLR dynamical modeling used in this work. Galaxy name and redshift are found in Columns 1 and 2, respectively. Column 3 refers to the published \texttt{CARAMEL} results. The sample has increased from 28 \citepalias[see][]{2023ApJ...948...95V} to 38, with the additional sources of \citetalias{2023ApJ...944...29B}, \citetalias{2023ApJ...959...25B}, and \citetalias{Wang_2026}.} 
\label{tab: geometry}
\end{deluxetable}

\begin{figure*}[h]
\centering
\begin{minipage}[b]{1.0\textwidth}
\centering
\includegraphics[width=1.0\textwidth]{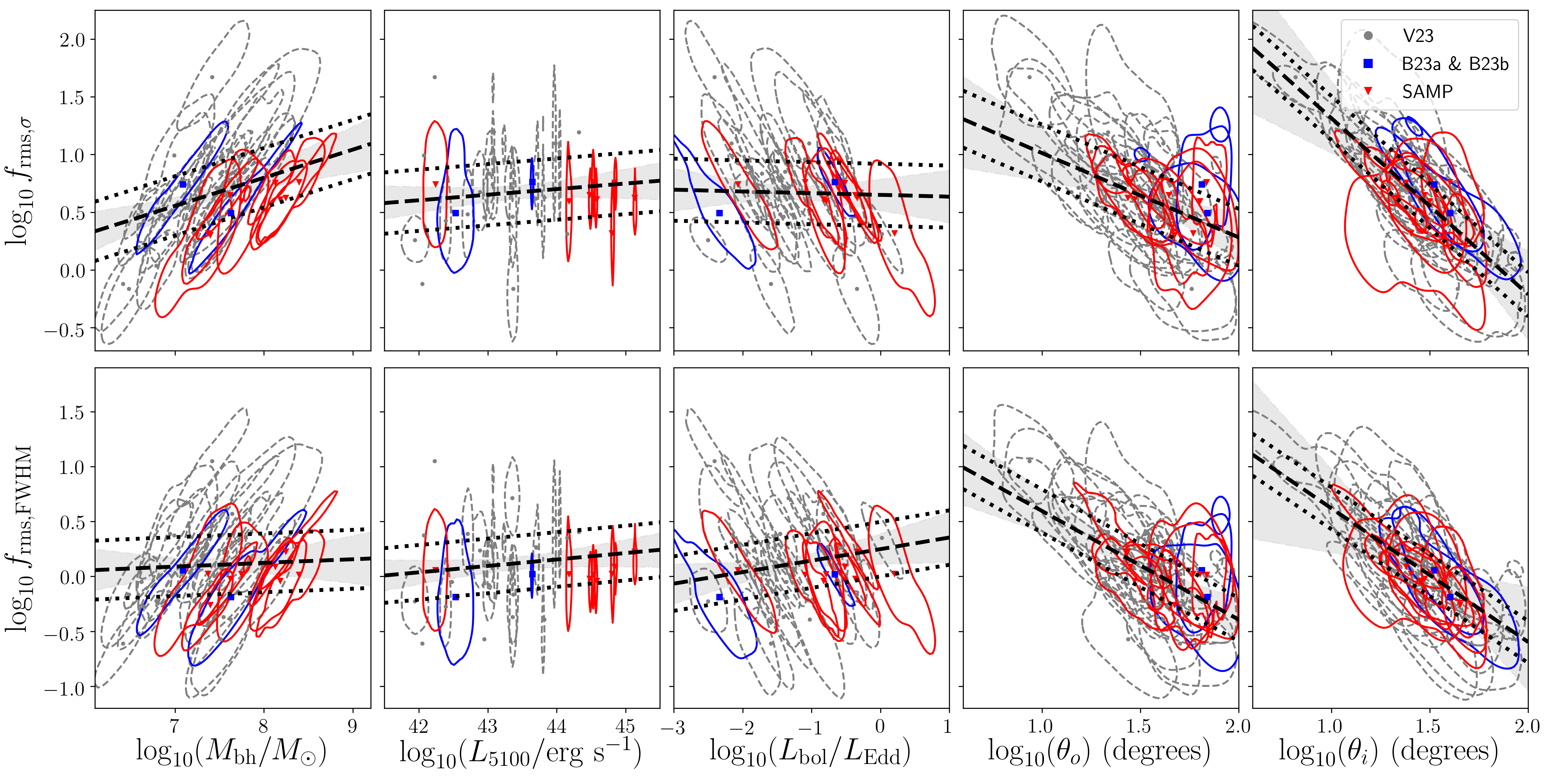}
\caption{Correlations between \logfrmssigma\ and \logfrmsfwhm\ with select BLR/AGN parameters. From left to right: \mbh, optical luminosity, Eddington ratio, \Hb-emitting BLR opening angle (disk thickness), and \Hb-emitting BLR inclination angle. The dashed black lines and gray shaded regions give the median and 68\% confidence intervals of the linear regression. Dotted lines are offset above and below the dashed line by the median value of the intrinsic scatter. Grey points represent the  sample used in \citetalias{2023ApJ...948...95V}. This work includes the addition of 10 new sources -- \citetalias{2023ApJ...959...25B} and \citetalias{2023ApJ...944...29B} are indicated by blue squares, and SAMP \citepalias{Wang_2026} are shown in red triangles. The contours surrounding each point show the 68\% confidence regions of the 2D posterior PDFs for each AGN.}
 \label{fig:f_correlations_rms_potato}
\end{minipage}
\begin{minipage}[b]{1.0\textwidth}
\centering
\includegraphics[width=\textwidth]{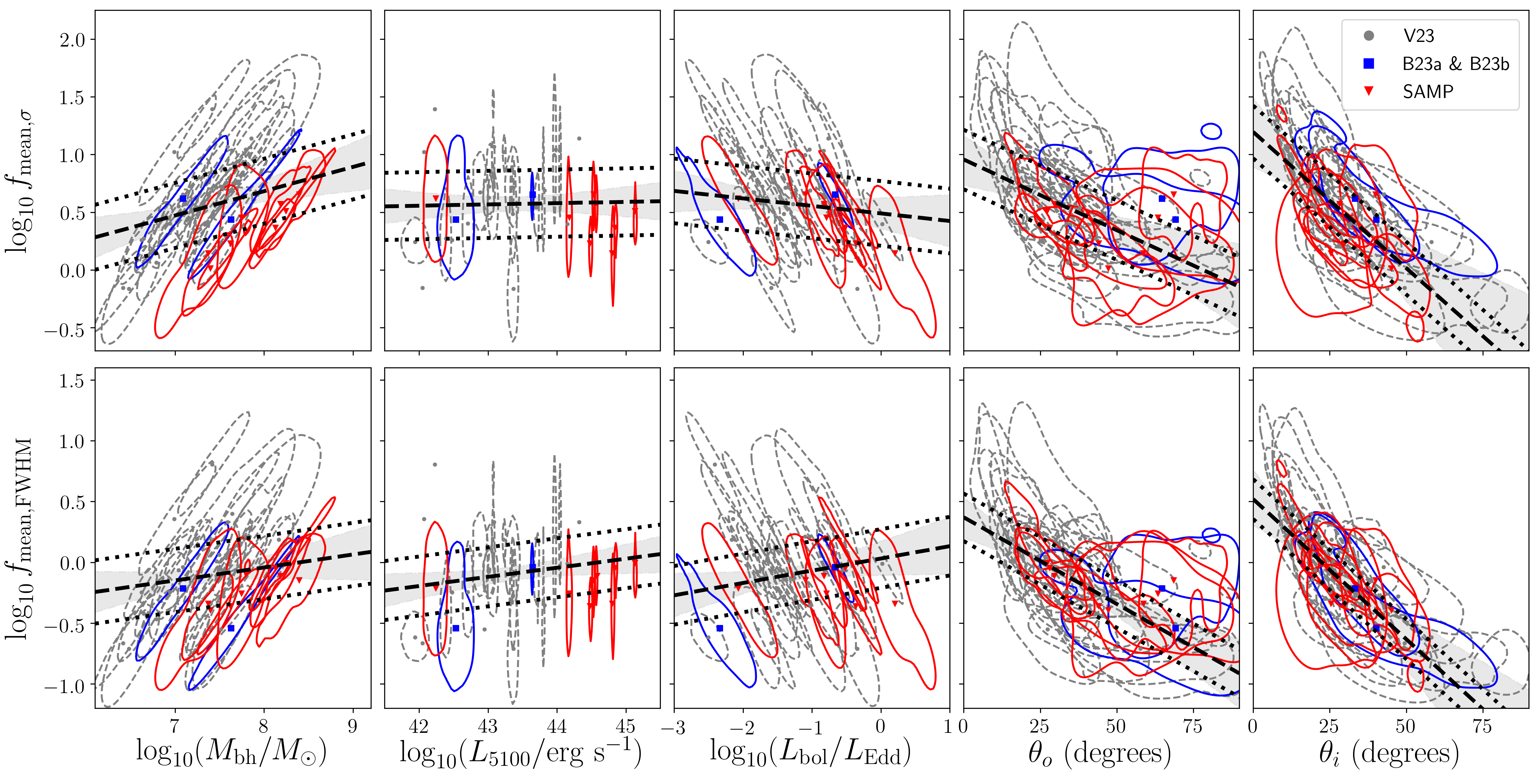}
\caption{Same as Figure \ref{fig:f_correlations_rms}, for correlations between \logfmeansigma\ and \logfmeanfwhm\ with select BLR/AGN parameters. The contours surrounding each point show the 68\% confidence regions of the 2D posterior PDFs for each AGN.}
\label{fig:f_correlations_mean_potato}
\end{minipage}
\end{figure*}

\end{appendix}
\bibliographystyle{apj}
\bibliography{references}

\end{document}